\documentclass[lettersize,journal]{IEEEtran}
\usepackage{amsmath,amsfonts}
\usepackage{algorithmic}
\usepackage{algorithm}
\usepackage{array}
\usepackage[caption=false,font=normalsize,labelfont=sf,textfont=sf]{subfig}
\usepackage{textcomp}
\usepackage{stfloats}
\usepackage{url}
\usepackage{verbatim}
\usepackage{graphicx}
\usepackage{cite}
\begin{document}

\title{Orthogonal Mode Decomposition for Finite Discrete Signals}

\author{
    \fontsize{12pt}{18pt}\selectfont{Ning Li}
    \\
    \fontsize{10pt}{15.75pt} \selectfont{ (Nanjing Institute of Technology)}    \\
    \fontsize{10pt}{15.75pt} zdhxln@njit.edu.cn \\
        \and 
    \fontsize{12pt}{18pt} \selectfont{Lezhi Li} \\
    \fontsize{10pt}{15.75pt} \selectfont{ (ML Collective)} \\
    \fontsize{10pt}{15.75pt} lli2@gsd.harvard.edu\\
}

\date{}

\maketitle

\begin{abstract}
In this paper, an orthogonal mode decomposition method is proposed to decompose finite-length real signals on the real and imaginary axes of the complex plane. The interpolation function space of finite length discrete signal is constructed, and the relationship between the dimensionality of the interpolation function space and its subspaces and the bandwidth of the interpolation function is analyzed. It is proven that the intrinsic mode is a narrow-band signal whose intrinsic instantaneous frequency is always positive (or always negative). Thus, the mode decomposition is transformed into the orthogonal projection of the interpolation function space to its low-frequency or narrow-band subspace. Unlike existing mode decomposition methods, the orthogonal mode decomposition is a local time-frequency domain algorithm. Each operation extracts a specific mode. The global decomposition results obtained under the precise definition of modes have uniqueness and orthogonality. The computational complexity of the orthogonal mode decomposition method is also much smaller than that of the existing mode decomposition methods.
\end{abstract}



\section{introduction}
\label{introduction}
\IEEEPARstart{T}
he time-frequency domain analysis method (TF) is useful in processing non-stationary time series signals. Early time-frequency domain analysis methods include the short-time Fourier transform and the wavelet transform\cite{huang1998empirical}. In 1998, Huang et al.\cite{huang1998empirical} proposed Empirical Mode Decomposition (EMD), which recursively extracts the modes of a signal from the high-frequency component to the low-frequency component. The specific process is finding all the maximum and minimum points of the signal and connecting all the maximum and minimum points to form the upper and lower envelopes through interpolation. The mean value of the upper and lower envelopes is removed as the low-frequency component, and the remaining part is an intrinsic mode function (IMF). Recursively running the above process on the low-frequency components, all the remaining modes can be extracted in turn. The empirical mode decomposition method has been widely used in time-frequency domain analysis. However, because of the lack of strict mathematical principles, the EMD has the unavoidable drawbacks of mixing among different modes. Moreover, EMD is a recursive global mode decomposition method, which cannot only extract a specific mode without considering other modes.

In 2014, Dragomiretskiy et al. proposed the Variational Mode Decomposition (VMD) \cite{dragomiretskiy2013variational} method. The VMD is an adaptive mode decomposition. It is not recursive. To use VMD, one should manually specify the number of modes contained in the signal. Using the variational equation, the center frequency and bandwidth of the modes can be found adaptively in the frequency domain. Mode decomposition can be achieved. The VMD method has a strict mathematical theoretical foundation, and the defect of mode mixing is also overcome to some extent. However, determining the number of mode decomposition layers in VMD is crucial. Artificially specifying the number of modes brings new drawbacks. When the specified number of modes is too large, the actual modes will be cracked. Otherwise, if the specified number of modes is too small, some modes may be stacked together and cannot be decomposed. Although the VMD algorithm is not recursive, it is also a global mode decomposition method that cannot only extract a specific mode without considering other modes. 

All modes in a real signal $\boldsymbol{u}$ form a mode set. The mode set is the intrinsic characteristic of the real signal $\boldsymbol{u}$. It should be unique, preferably orthogonal. However, the mode set obtained using EMD and VMD is not unique and does not satisfy orthogonality.

This article introduces a new mode decomposition algorithm for finite discrete real signals. The mode set obtained by this method is unique and orthogonal, and we call it the intrinsic mode set. We name our method the Orthogonal Mode Decomposition method. Compared with the existing methods, the mode decomposition method proposed in this paper has the following advantages:

\begin{enumerate}
  \item The mode decomposition method presented in this article is based on the orthogonal projection algorithm, where all modes contained in the original signal are orthogonal.
  \item Accompanied by the orthogonality is the uniqueness of mode decomposition. The precise definition of modes ensures that all modes have clear center positions and widths in the frequency domain, and no overlap among modes. 
  \item The mode decomposition algorithm presented in this article is based on clear mathematical principles.
  \item Unlike EMD and VMD, the orthogonal mode decomposition method is not based on global decomposition. It can focus on and extract specific modes of interest without extracting all modes. This feature reduces computational complexity.
  \item The examples show that the accuracy of mode decomposition can be maintained even at both ends of the time segment. The boundary effect inherent in the traditional mode decomposition method is overcome.
\end{enumerate}

This article will unfold in the following way. First, we will discuss the definition of mode in Section \ref {sec:problem_formulation}, which is the theoretical basis of the decomposition method proposed in this article. Next, in section \ref {sec:ortho_mode_extraction}, the principle and method of extracting modes based on orthogonal projection will be introduced. The mode set obtained using this method has uniqueness and orthogonality. The specific algorithm for calculating the intrinsic phase function and the instantaneous frequency of the signal will be detailed in Section \ref {sec:inherent_frequency}, which can determine the boundary of the mode in the frequency domain. Section \ref {sec:search_mode} discusses the method of searching for the center frequency of the mode. Section \ref {sec:low_freq} discusses using resampling techniques to remove low-frequency non-oscillatory components from the original signal. The above contents are explained with examples. In Section \ref {sec:comparison}, we compare the method proposed in this article with previous methods and discuss its superiority. The examples used in this article all have Python 3.8 programs that can be obtained through the author's email.

\section{What is a Mode?}
\label{sec:problem_formulation}

The unique mode set is based on a precise definition of modes\cite{rilling2007one}. This section discusses the definition of mode, this is the basis of the mode decomposition method introduced in this article.

Since Huang et al.\cite{huang1998empirical} proposed Empirical Mode Decomposition (EMD) in 1998, the method of mode composition has been widely applied in various domains such as electrocardiogram analysis, earthquake research, climate research, mechanical fault diagnosis, economic statistical analysis, etc. However, the definition of mode is constantly a matter of debate. The original definition of mode was: 1) The number of extreme points is equal to, or differs from the number of zero crossing points by at most 1, throughout the entire time course; 2)Two envelope lines formed by all maximum and minimum values have a mean of zero, at any point.

In subsequent work (especially in Variational Mode Decomposition), there were minor changes to the definition of modes. It is believed that a mode $m_ i (t) $ is a frequency modulation and amplitude modulation function  \cite{dragomiretskiy2013variational,daubechies2011synchrosqueezed,hussain2002adaptive}:
$$m_i(t)=A_i(t)cos(\Phi_i(t)),i=1,\ldots, M$$

Among them, $A_i (t) $and $\Phi_i (t) $ are functions that change slowly, $A_i (t) $ is the amplitude of the mode, $\Phi_i (t) $ is its phase.\cite{dragomiretskiy2013variational, hussain2002adaptive}.

Some literature has discussed the above two definitions and believes that the latter definition of mode is more rigorous than the former\cite{dragomiretskiy2013variational, daubechies2011synchrosqueezed}. However, the two definitions mentioned above are descriptive and have some uncertainty. From a practical application point of view, the modes must be narrow-band functions. That is to say, the Fourier bandwidth of the mode is limited.  However, how to limit the bandwidth of modes has not been discussed in EMD-related theories. In the EMD, the uncertainty of the mode bandwidth causes aliasing between different modes \cite{dragomiretskiy2013variational, maji2016empirical}. VMD algorithm needs to artificially specify the number of modes, and obtain the specified number of modes by iteratively solving the variational equation in the frequency domain. The modes obtained by VMD are generally frequency and amplitude modulation functions. The second definition of mode is derived from this. VMD solves the problem of frequency band overlap between different modes to some extent, but the VMD cannot accurately specify the spectral position of the mode of interest \cite {guo2021generalized}. The above two definitions of modes lack the limit on the bandwidth of the mode itself and the description of the characteristics associated with the bandwidth, so the decomposed set of modes does not have uniqueness.

Although the results obtained by previous mode decomposition methods are not unique, these methods still play an important role in processing non-stationary signals\cite{lerga2011efficient}. Because the harmonic components in the mode are relatively simple, the mode can be regarded as a stationary signal to a certain extent. The Hilbert transform is used to obtain the phase function of the mode, which can remain monotonic throughout the entire time range. The monotonicity of the phase function of the modes makes the instantaneous frequency meaningful, which cannot be done for the signals with complex frequency components. The so-called mode is formed by gathering the maximum number of harmonic components under the premise of the monotonicity of its phase function. Therefore, calculating the phase function and instantaneous frequency is key to accurately extracting modes.
\section{Extracting Modes Based on Orthogonal Projection Algorithm}
\label{sec:ortho_mode_extraction}
\vspace{-0.2cm}
\subsection{The interpolation function of the finite discrete signal} 

The actual discrete real signals are all of finite length, and the discussion of the finite-length discrete time domain real signals (time series signals) has important theoretical and practical significance.

The finite discrete real signals$\hspace{2mm} \boldsymbol{u}=[u(-l),u(-l+1),\hdots,u(0),\hdots,u(l-1),u(l)]^T$, The length of $\boldsymbol{u}\hspace{1mm} $ is $2l+1$. Let $n = 2l+1$ then $\boldsymbol{u} \in R^n $, $R^n$ is an n-dimensional vector space in the real number field. The sampling period of the discrete real signals$\boldsymbol{u}$ is $\Delta$, and the entire time segment is $T=(n-1)\Delta$. Constructing the interpolation function $\Psi_u (t)$ using the series $\boldsymbol{u}$:

\begin{equation}
\Psi_u(t) = \sum_{k=-l}^{l}{u(k) \frac{\sin \Omega_{\Delta}(t - k\Delta) }{\Omega_{\Delta}(t - k\Delta)}}\label{psi} 
\end{equation}

where $t \in (-\infty, \infty)$, $\Omega_{\Delta} = \pi / \Delta$. the following expression holds truly:
\begin{equation}
  \Psi_{u}(k \Delta) =
    \begin{cases}
      u(k) & \text{if $k = -l, (-l+1), \hdots, 0, \hdots, (l-1), l$ }\\
      0 & \text{if $k > l$ or $k < -l$}.
    \end{cases}   \nonumber     
\end{equation}

\begin{figure}[ht]
\centering
\includegraphics[width=7cm,height=3.2cm]{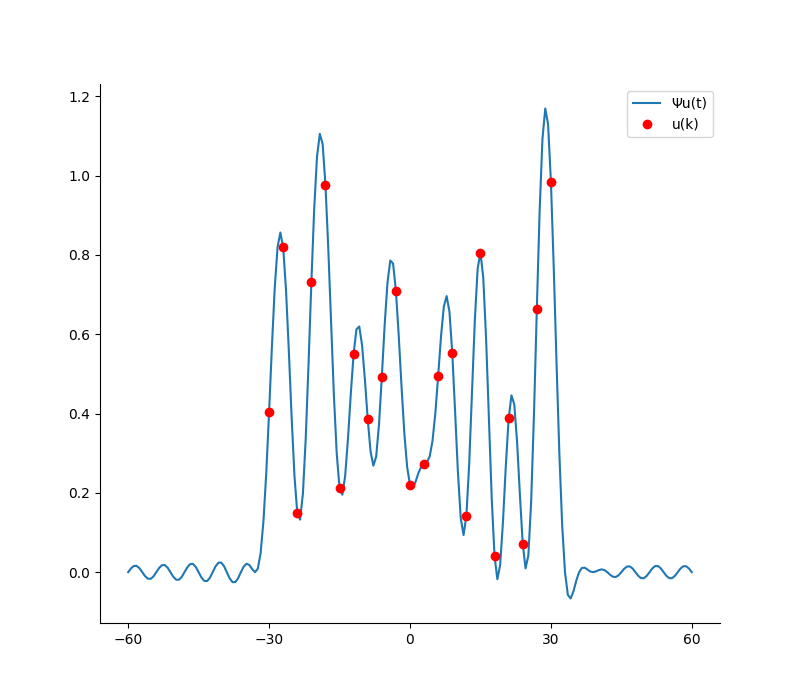}
\caption{\label{fig:fig_0}  The interpolation function for the finite length discrete real signals}
\end{figure}

$\Psi_u (t)$ is the function defined on the infinite interval$(-\infty, \infty)$, but its main support range is $[-l\Delta,l\Delta]$. The main support range is the principal value interval of $\Psi_u (t)$. Beyond the principal value interval, $\Psi_u(t)$ tends to zero.  In the time domain, the energy of $\Psi_u (t)$ is concentrated in the principal value interval. Figure \ref{fig:fig_0} shows an example of the interpolation function for finite discrete real signals.

\textbf{Example0:}  In this example, the length of the random discrete signal $\boldsymbol{u} $ is 21, it is distributed in the interval $[-30,30]$. The interpolation function $\Psi_{u}(t)$ is generated according to the formula \ref{psi}, and the principal value interval of the interpolation function is also $[-30,30]$.

The Fourier transform of $\Psi_u(t)$ is:
\begin{align}
F[\Psi_u(t)] & = \int_{-\infty}^{\infty} \Psi_u(t) e^{-j\omega t} \,dt \nonumber\\
&= \int_{-\infty}^{\infty} \sum_{k=-l}^{l}{u(k) \frac{\sin \Omega_{\Delta}(t - k\Delta) }{\Omega_{\Delta}(t - k\Delta)}} e^{-j\omega t} \,dt \nonumber \\
             & = \sum_{k=-l}^{l}{u(k) \int_{-\infty}^{\infty} \frac{\sin \Omega_{\Delta}t'}{\Omega_{\Delta}t'}} e^{-j\omega (t' + k\Delta)} \,dt' \nonumber \\ 
             & = \sum_{k=-l}^{l}{u(k) e^{-j\omega k\Delta}} \int_{-\infty}^{\infty} \frac{\sin \Omega_{\Delta}t}
              {\Omega_{\Delta}t}e^{-j\omega t} \,dt\nonumber \\
              &= U_u(\omega)U_{\Omega_{\Delta}}(\omega)
              \nonumber
\end{align} 

Among them:
\begin{flalign}
& U_{\Omega_{\Delta}}(\omega) = \int_{-\infty}^{\infty} \frac{\sin \Omega_{\Delta}t} {\Omega_{\Delta}t}e^{-j\omega t} \,dt = 
    \begin{cases}
      \Delta & |\omega| <= \Omega_{\Delta} \\
      0      & |\omega| > \Omega_{\Delta} 
    \end{cases} 
    \nonumber
\\
& U_u(\omega) = \sum_{k=-l}^{l}{u(k) e^{-j\omega k\Delta}} 
\nonumber
\end{flalign}

\begin{figure}[ht]
\centering
\includegraphics[width=4cm,height=2.9cm]{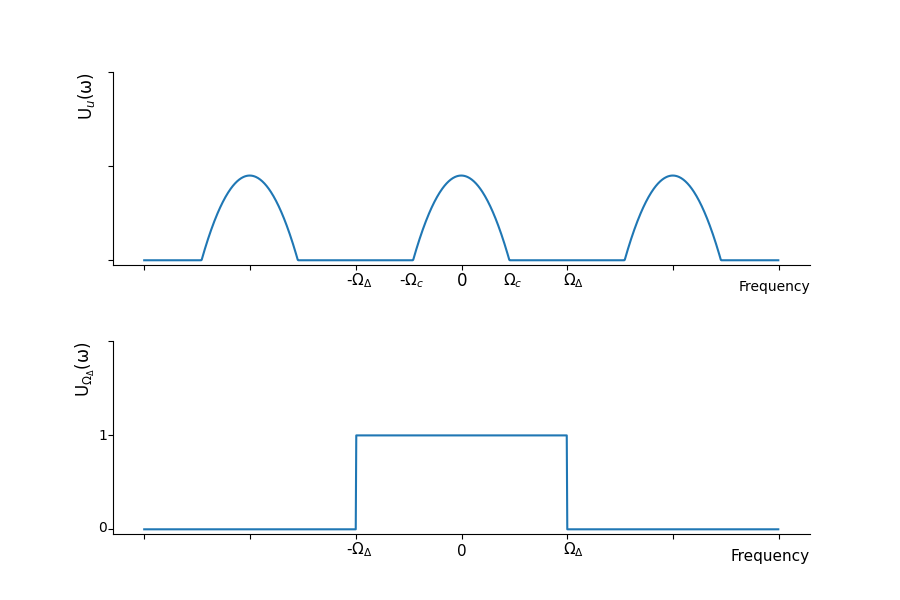}
\caption{\label{fig:fig1}$U_u (\omega)$ and $U_{\Omega_\Delta} (\omega)$ }
\end{figure}

As shown in Figure$\hspace{1mm}$\ref{fig:fig1}, $U_u (\omega)$ is the frequency domain function with the period of $2\Omega_\Delta$.
$U_{\Omega_\Delta} (\omega)$ is Frequency domain window function.
The bandwidth of $\Psi_u(t)$ is $ \hspace{1mm} \Omega_c$, which is the cutoff frequency of the main lobe of $U_u (\omega)$. 
$\Omega_c<\Omega_\Delta$ and $\Omega_\Delta=\frac{\pi}{\Delta}$.
The bandwidth of the finite discrete signal $\boldsymbol{u}$ is the same as that of the continuous function $\Psi_u(t)$. In the frequency domain, the continuous function $\Psi_u (t)$'s energy is concentrated in the principal value interval.

Label the set of all interpolation functions $\Psi_u(t)$ as $\boldsymbol{IFS_{(n, \Delta)}}$. The bandwidth of all the interpolation functions$\Psi_u(t)$ in $\boldsymbol{IFS_{(n, \Delta)}}$ are smaller than $\Omega_\Delta$, $\Omega_\Delta=\frac{\pi}{\Delta}$. Label $n$ as the number of sampling points for the original signal$ \hspace{1 mm}\boldsymbol{u}$, $n=2*l+1$. The frequency domain resolution of $\Psi_u(t)$ is $\epsilon$, $\epsilon = \frac{\pi}{l\Delta} $. Due to the one-to-one correspondence between $\Psi_u(t)\in\boldsymbol{IFS_{(n, \Delta)}}$ and $\boldsymbol{u}\in R^n$, it is known that $\boldsymbol{IFS_{(n, \Delta)}}$ and $R^n$ are isomorphic.
Consider the function set:

\begin{flalign}
\label{eq:ortho_basis_whole}
&\{1,sin(\frac{\pi t}{l\Delta}),cos(\frac{\pi t}{l\Delta}),\hdots,sin(\frac{i \pi t}{l\Delta}),cos(\frac{i \pi t}{l\Delta}), \hdots,\nonumber\\
&sin(\frac{\pi t}{\Delta}),cos(\frac{\pi t}{\Delta})\}
\end{flalign}

The function set(\ref{eq:ortho_basis_whole}) can be used as an orthogonal basis of $\boldsymbol{IFS_{(n, \Delta)}}$.
The set of orthogonal basis functions is made up of n single-frequency functions. $\forall \Psi_u(t) \in \boldsymbol{IFS_{(n, \Delta)}}$, $\Psi_u(t)$ can be uniquely expressed by the linear combination of the basis functions in set(\ref{eq:ortho_basis_whole}). But this is not the mode decomposition we want. In actual signal analysis, some modes that contain multiple frequency components often correspond to the specific characteristics of the actual physical system. The modes should not be split into single-frequency components.

\subsection{Mode extraction method based on orthogonal projection}
\label{subsection:mode_extraction}

Considering frequency domain resolution $\epsilon$, $\epsilon = \frac{\pi}{l\Delta} $. Let $\hspace{1mm}\Omega_{c1} = \frac{i\pi}{l\Delta}$,  $\Omega_{c2} = \frac{m\pi}{l\Delta}$. $i$ and $m$ are positive integers, and $0<i<m<l$. All narrow band functions with $\Omega_{c1}$ and $\Omega_{c2}$ as frequency band boundaries form the subspace $\boldsymbol{IFS_{(n, \Delta, \Omega_{c2} - \Omega_{c1})}}$,
which is the subspace of $\boldsymbol{IFS_{(n, \Delta)}}$.

The subset of the the function set(\ref{eq:ortho_basis_whole}) forms the orthogonal basis of $\boldsymbol{IFS_{(n, \Delta, \Omega_{c2} - \Omega_{c1})}}$:
\begin{equation}
\label{eq:ortho_basis}
    \{\sin(\frac{i\pi t}{l\Delta}), cos(\frac{i\pi t}{l\Delta}), \hdots, \sin(\frac{m \pi t}{l\Delta}), cos(\frac{m \pi t}{l\Delta})\}
\end{equation}

Let $\hspace{1mm}p = m-i+1$. The subset (\ref{eq:ortho_basis}) contains $2p$ single-frequency functions. 

Let $\boldsymbol{t}=[-l,-l+1,\hdots,0,\hdots,l-1,l]^T\Delta$. $\boldsymbol{t}$ is a time vector composed of all sampling points. To substitute the time vector $\boldsymbol{t}$ into every function in the subset(\ref{eq:ortho_basis}), the $2p$ column vectors can be obtained. Every column vector is $n$ dimensional. The two column vectors corresponding to the frequency point $v$ are:
\begin{align}
\label{eq:v_frequent_point}
&[sin(-v\pi), \hdots, sin(-\frac{v\pi}{l}), 0, sin(\frac{v\pi}{l}),\hdots, sin(v\pi)]^T\\
&[cos(-v\pi), \hdots, cos(-\frac{v\pi}{l}), 1,  cos(\frac{v\pi}{l}),\hdots, cos(v\pi)]^T
\end{align}

Among them $v = i, i+1, ..., m-1, m$.

 These column vectors can form the $n \times 2p$ order matrix $\Gamma$. Thus, the orthogonal projection matrix $\Gamma(\Gamma^T \Gamma)^{-1}\Gamma^T$ can be obtained. For all $\Psi_u (t) \in \boldsymbol{IFS_{(n, \Delta)}}$, the orthogonal projection matrix can be used to find its mode in the subspace $\boldsymbol{IFS_{(n, \Delta, \Omega_{c2} - \Omega_{c1})}}$:

\begin{equation}
\label{eq:mode_local}
   \Psi_u^{[\Omega_{c1}, \Omega_{c2}]} = \Gamma(\Gamma^T \Gamma)^{-1}\Gamma^T \boldsymbol{u} 
\end{equation}

The mode extraction method based on the orthogonal projection given in Equation (\ref{eq:mode_local}) is a local method that only needs to extract any mode of interest without considering other modes.

However, the problem has not been completely solved at this point. The next questions are: How to determine the frequency band interval $[\Omega_{c1}, \Omega_{c2}]$? How do we know if this frequency band interval covers a complete mode?

In the following paragraph, we shall discuss combining the largest number of frequency components while keeping the intrinsic phase functions monotonous. The sum of the largest number of frequency components is a mode in the precise sense, as long as the intrinsic phase function of the sum is monotonic.  Calculating the intrinsic phase function is the key to solving this problem. For this purpose, we will introduce the specific methods for calculating the intrinsic phase function and instantaneous frequency.

\section{The intrinsic phase function and instantaneous frequency of the finite-length discrete signals}
\label{sec:inherent_frequency}

In traditional spectrum analysis based on the Fourier transform, frequency is the time-independent quantity and essentially the overall characteristic of the finite-length signal. Fourier frequency is effective for describing the characteristics of stationary signals. Nevertheless, the Fourier frequency is not effective in non-stationary signals. That is because non-stationary signals change at any time randomly. For researching the local characteristics of the signals, the concepts of phase function and instantaneous frequency are introduced as follows\cite{boashash1992estimating,huang2009instantaneous}:

For consistency of symbolic expression, we also label the general time domain signal as $\Psi_u (t)$.The analytic signal about $\Psi_u (t)$ is:
$$z(t)=\Psi_u (t)+jH\{\Psi_u (t) \}=a(t) e^{j\Theta(t)} $$

$H\{\Psi_u (t)\}$ is the Hilbert transform of $\Psi_u (t)$. If $\Psi_u (t)$ is a single-frequency function, $H\{\Psi_u(t)\}$ is like a 90 degree phase shifter. $H\{\Psi_u(t)\}$  produces a negative $90^\circ$ phase shift on the positive frequency component and a positive $90^\circ$ phase shift on the negative frequency component.
$a(t)$ is the time-varying amplitude of the signal and $\theta(t)$ is the phase function of the signal. The derivative of $\theta(t)$ is $\theta'(t) $, and  $\theta'(t) $is actually the instantaneous frequency\cite{boashash2015time,huang1998empirical,dragomiretskiy2013variational}.

Using instantaneous frequency is convenient for analyzing non-stationary signals, but its calculation is difficult \cite{hussain2002adaptive,huang2009instantaneous}.
 Reference \cite{huang2009instantaneous} gives a direct integration method for obtaining the phase shift of $-90^\circ$ of the original real signal $\Psi_u (t)$ so that the analytic signal z(t) can be obtained without the Hilbert transform, thus the instantaneous frequency of the real signal $\Psi_u (t)$ can be approximately calculated.
 The exploration of the true physical meaning and the calculation of instantaneous frequency has been ongoing, and there is still significant research  
 space\cite{boashash2015time,huang2009instantaneous}

This paper presents a method for obtaining the intrinsic phase function and the instantaneous frequency of a real signal based on the parity decomposition. This method is simple in calculation and clear in physical meaning. In some typical examples, the results obtained using this method are the same as those obtained using the Hilbert transform.

For $ \boldsymbol{u} = [u(-l), u(-l+1),\hdots, u(0),\hdots, u(l-1), u(l)]^T, \boldsymbol{u} \in R^n $, let $ \boldsymbol{u_{inv}} $  be in reverse order of $\boldsymbol{u}$, that is 
$ \boldsymbol{u_{inv}} = [u(l), u(l-1),\hdots, u(0), \hdots, u(-l+1), u(-l)]^T$, and let:
\begin{align} \label{orth_decomp}
\begin{split}
    \boldsymbol{u_e} & = 0.5(\boldsymbol{u} + \boldsymbol{u_{inv}}) \\
    \boldsymbol{u_o} & = 0.5(\boldsymbol{u} - \boldsymbol{u_{inv}}) \\
    \boldsymbol{u} & = \boldsymbol{u_e} + \boldsymbol{u_o}
\end{split}
\end{align}

This decomposition is called the parity decomposition of the discrete signal $\boldsymbol{u}$, where $\boldsymbol{u_e}$ is an even component and $\boldsymbol{u_o}$ is an odd  component, and there are:
\begin{align} \label{orth_decomp1}
& \begin{cases}
    u_e(-k) = u_e(k) \\
    u_o(-k) = -u_o(k)
  \end{cases}
& k = 1, 2,\ldots,l 
& \text{      }
& \begin{cases}
  u_e(0) = u(0) \\
  u_o(0) = 0
\end{cases} \nonumber
\end{align}

Construct interpolation functions $ \Psi_{u}(t) $, $ \Psi_{u_e}(t) $, $ \Psi_{u_o}(t) $, for $ \boldsymbol{u}$, $\boldsymbol{u_e}$ and $ \boldsymbol{u_o}$ separately.
  Clearly, $ \Psi_{u}(t) = \Psi_{u_e}(t) + \Psi_{u_o}(t) $. $ \Psi_{u_e}(t) $ is an even function and $ \Psi_{u_o}(t) $ is an odd function. 
 We perform Fourier transform on $ \Psi_{u_e}(t) $ and $ \Psi_{u_o}(t) $ separately, yields:  
\begin{align}
    F[\Psi_{u_e}(t)]  &= [u(0) + \sum_{k = 1}^{l}{2u_e(k) \cos(\omega k \Delta)}] U_{\Omega_{\Delta}}(\omega), \nonumber \\
    F[\Psi_{u_o}(t)]  &= -j[\sum_{k = 1}^{l}{2u_o(k) \sin(\omega k \Delta)}] U_{\Omega_{\Delta}}(\omega) \nonumber
\end{align} 

The Fourier transform of $\Psi_{u_e}(t)$ is a real function denoted as $Fre(\omega)$. The Fourier transform of $\Psi_{u_o}(t)$ is a pure imaginary function denoted as -$jFim(\omega)$.
Therefore, $\boldsymbol{u_e}$ is also referred to as the even component of the original signal $\boldsymbol{u}$ and $\boldsymbol{u_o}$ is also referred to as the odd component of $\boldsymbol{u}$.  $Fre(\omega)$ is referred to as the real part spectrum of $\boldsymbol{u}$ and $Fim(\omega)$ is referred to as the imaginary part spectrum of $\boldsymbol{u}$.  Together, they can be written as:
$$F[\Psi_u(t)] = Fre(\omega) - jFim(\omega)$$

It can be considered that $\Psi_{u_o}(t)$ has a phase difference of $-90^\circ$ relative to $\Psi_{u_e}(t)$. If $\Psi_{u_e}(t)$ and $\Psi_{u_o}(t)$ are not zero, then the amplitude $A(t)$ and phase $\varphi(t)$ of $\Psi_{u}(t)$ are determined as follows:
\begin{align}
    A(t) &=\sqrt{\Psi_{u_e}^2 (t)+\Psi_{u_o}^2 (t)} \nonumber \\
    \varphi(t) &= -\tan^{-1}(\frac{\Psi_{u_o}}{\Psi_{u_e}}) \nonumber
\end{align} 

The instantaneous frequency of $\Psi_{u}(t)$ is defined as:
$$\omega(t) = \frac{d\varphi(t)}{dt} = \frac{\Psi_{u_e}' (t) \Psi_{u_o} (t)-\Psi_{u_o}' (t) \Psi_{u_e} (t)}{A^2 (t)} $$
 
If $\Psi_{u} (t)$ is a pure even function, $\Psi_{u_o} (t) = 0$, $\Psi_{u} (t)=\Psi_{u_e} (t) = \sum_{i=0}^l{c_i \cos(\frac{i \pi t}{l\Delta})}$. Among them $\{c_i |i=0,1,\ldots ,l\}$ are the real coefficients. 
We can artificially define the odd function $f_q (t) = \sum_{i=1}^l{c_i \sin(\frac{i\pi t}{l\Delta})}$. The odd function $f_q (t)$ is orthogonal to $\Psi_{u_e} (t)$, which can be used instead of $\Psi_{u_o} (t)$ to calculate the phase function and the instantaneous frequency of $\Psi_{u} (t)$.

If $\Psi_{u} (t)$ is a pure odd function, $\Psi_{u_e} (t) = 0$, $\Psi_{u} (t)=\Psi_{u_o} (t) = \sum_{i=0}^l{s_i \sin(\frac{i \pi t}{l\Delta})}$. Among them $\{s_i |i=0,1,\dots,l\}$ are the real coefficients. 
We can artificially define the even function $f_d (t) = \sum_{i=1}^l {s_i \cos(\frac{i\pi t}{l\Delta})}$. The even function $f_d (t)$ is orthogonal to $\Psi_{u_o} (t)$, which can be used instead of $\Psi_{u_e} (t)$ to calculate the phase function and the instantaneous frequency of $\Psi_{u} (t)$.

According to the above definition of the instantaneous frequency $\omega(t)$ of a real signal $\boldsymbol{u}$, it is related to the even component $\Psi_{u_e} (t)$ and the odd component $\Psi_{u_o} (t)$ and the amplitude $A(t)$. Therefore, $\omega(t)$ is the intrinsic instantaneous frequency of the real signal $\boldsymbol{u}$. Some literature suggests that the instantaneous frequency of a real signal is meaningful only when it remains positive\cite{boashash2015time, huang1998empirical}. Here are some examples:

\vspace{1mm}

\textbf{Example1:} $\Psi_{u}(t)=sin(20t^3)+cos(20t^3), t \in [-1,1]$.

We sample $\Psi_{u}(t)$ at 100Hz.

\begin{figure}[ht] 
    \centering 
    \subfloat[]{        \includegraphics[width=3.1cm,height=3.1cm]{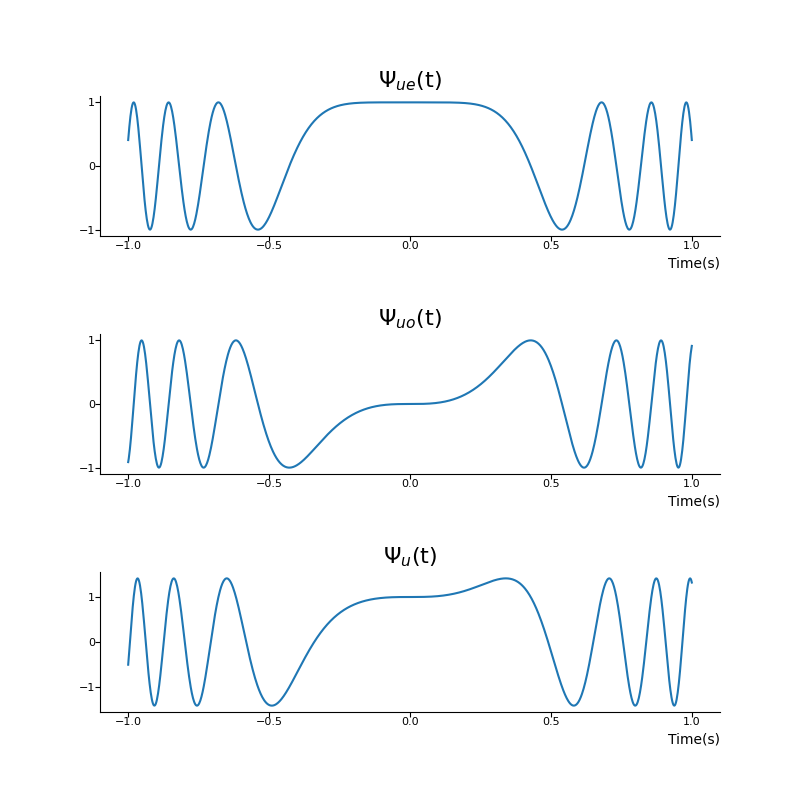}}%
    \hspace*{-3mm} 
    \subfloat[]{        \includegraphics[width=3.1cm,height=3.1cm]{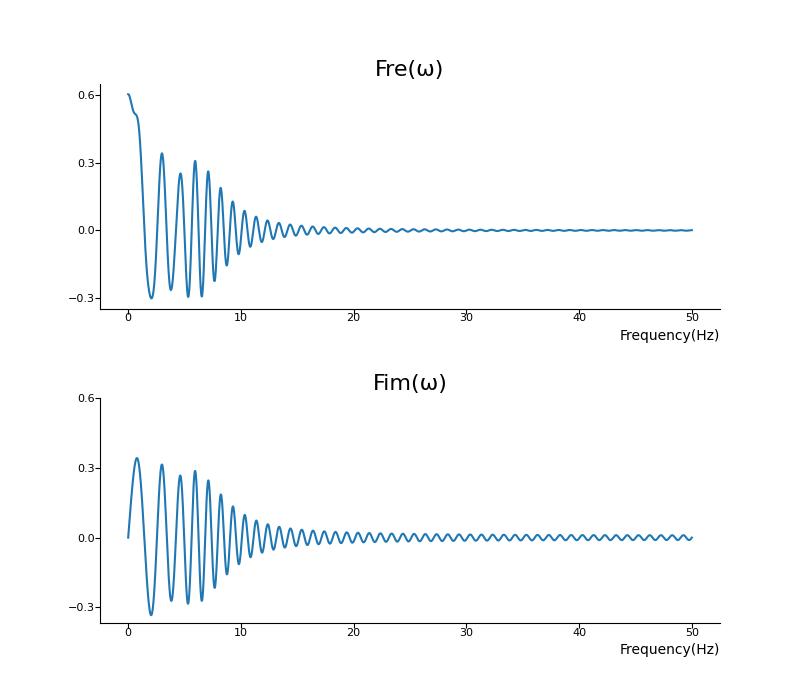}}%
    \hspace*{-3mm} 
    \subfloat[]{        \includegraphics[width=3.1cm,height=3.1cm]{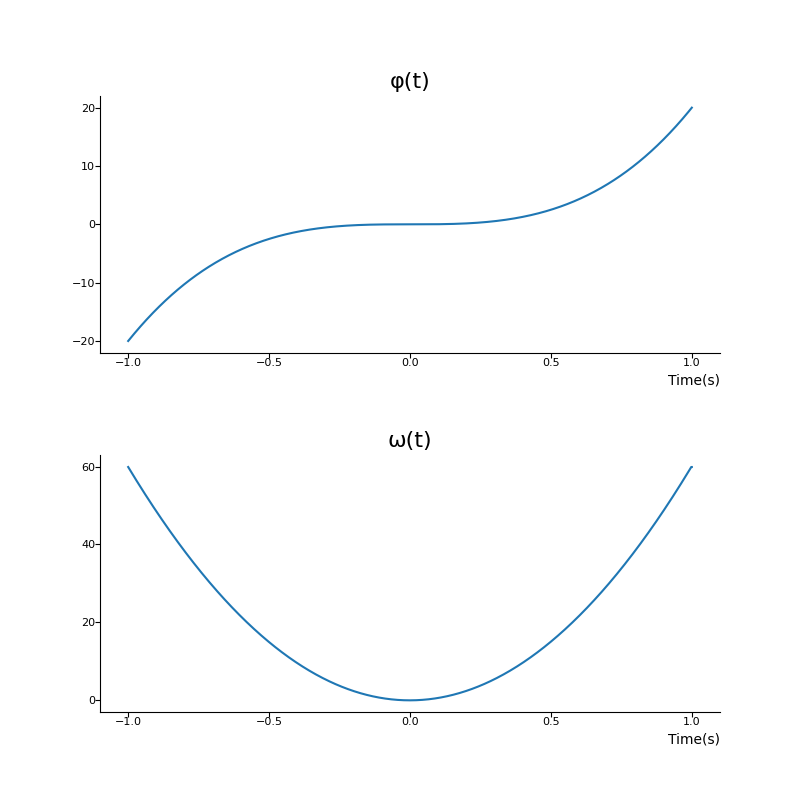}}%
    \caption {Example 1, $\omega(t)$ is always positive, and $\Psi_{u}(t)$ is a mode}\label{fig:fig_2} 
\end{figure}
In Example 1, the even component and the odd component of $\Psi_{u}(t)$ are shown in Figure \ref{fig:fig_2}(a). The real part spectrum and the imaginary part spectrum are shown in Figure \ref{fig:fig_2}(b). The phase function and intrinsic instantaneous frequency of $\Psi_{u}(t)$ are shown in Figure \ref{fig:fig_2}(c). The Fourier bandwidth of the signal is 0Hz $\sim$15Hz. Figure \ref{fig:fig_2}(c) shows that its intrinsic phase is a cubic function and monotonic. The intrinsic instantaneous frequency is a quadratic function and always positive, so $\Psi_{u}(t)$ is a mode.

\vspace{1mm}

\textbf{Example2:} $\Psi_{u}(t)=1.3sin(6\pi t)+sin(5\pi t)+1.5cos(6\pi t)+cos(5\pi t), t \in [-1,1]$.

We sample $\Psi_{u}(t)$ at 100Hz.

\begin{figure}[ht] 
    \centering 
    \subfloat[]{        \includegraphics[width=3.1cm,height=3.5cm]{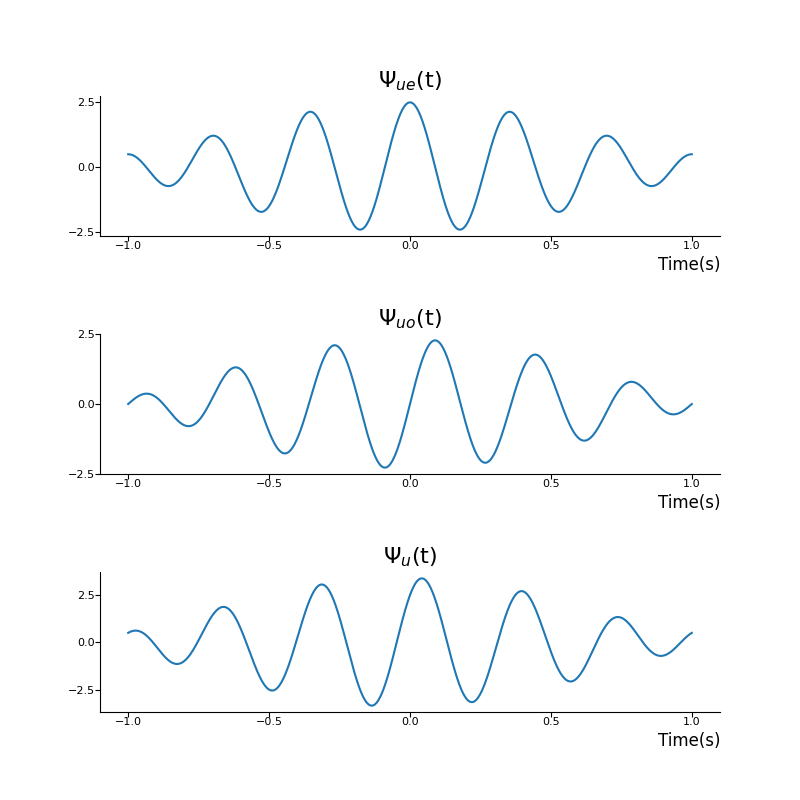}}      
    \hspace*{-3mm} 
    \subfloat[]{        \includegraphics[width=3.1cm,height=3.5cm]{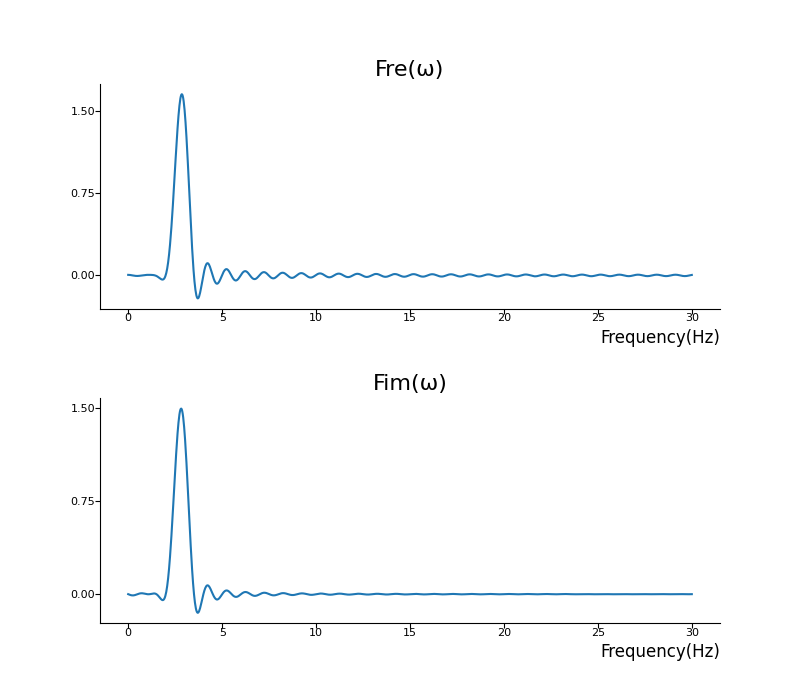}}      
    \hspace*{-3mm} 
    \subfloat[]{        \includegraphics[width=3.1cm,height=3.5cm]{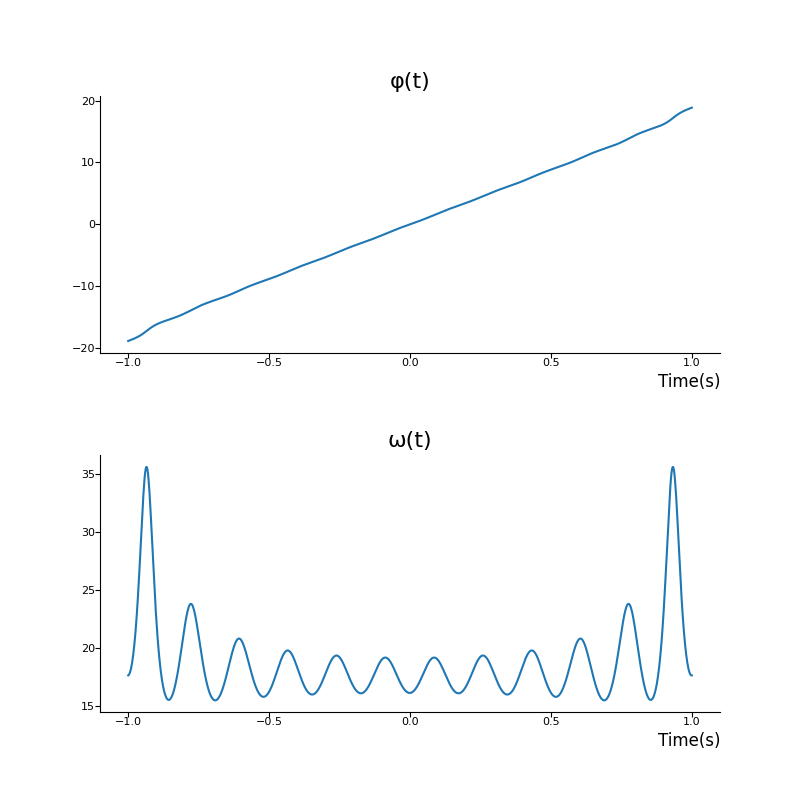}}%
    \caption{Example 2, both $Fre(\omega)$ and  $Fim(\omega)$ have only one principal lobe,$\Psi_{u}(t)$ is a mode}\label{fig:fig_3} 
\end{figure}

In Example 2, Figure \ref{fig:fig_3}(a) shows that $\Psi_{u}(t)$ and its real part component $\Psi_{u_e}(t)$ and imaginary part component $\Psi_{u_o}(t)$ are amplitude modulation functions. Figure \ref{fig:fig_3}(b) shows that both $Fre(\omega)$ and $Fim(\omega)$ have only one principal lobe. Figure \ref{fig:fig_3}(c) shows that the phase function $\varphi(t)$ is monotonic and the intrinsic instantaneous frequency $\omega(t)$ is always positive. $\Psi_{u}(t)$ can be considered as a mode.

\vspace{1mm}

\textbf{Example3:} $\Psi_{u}(t)=\sum_{f=1}^{87}{[sin(f\pi t)+cos(f\pi t)]} , t \in [-1,1]$.

We sample $\Psi_{u}(t)$ at 100Hz.

\begin{figure}[ht] 
    \centering
    \subfloat[]{        \includegraphics[width=3.1cm,height=3.5cm]{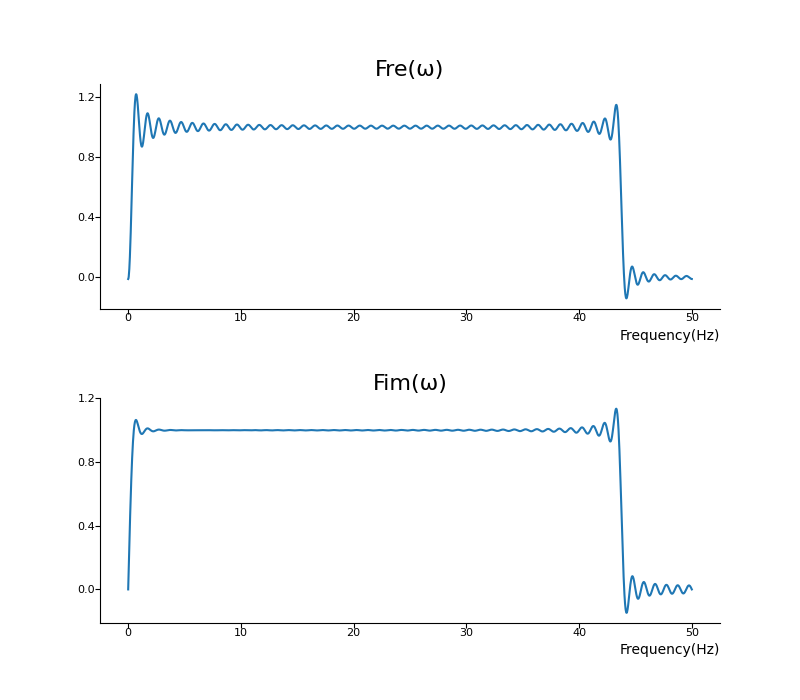}}  
    \hspace*{-3mm} 
    \subfloat[]{        \includegraphics[width=3.1cm,height=3.5cm]{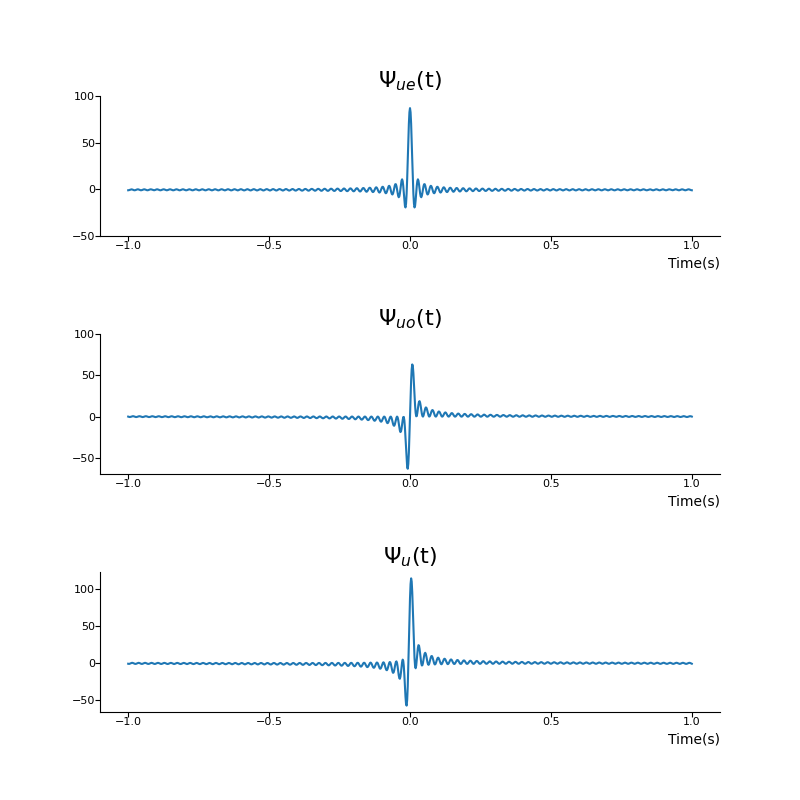}}%
    \hspace*{-3mm} 
    \subfloat[]{        \includegraphics[width=3.1cm,height=3.5cm]{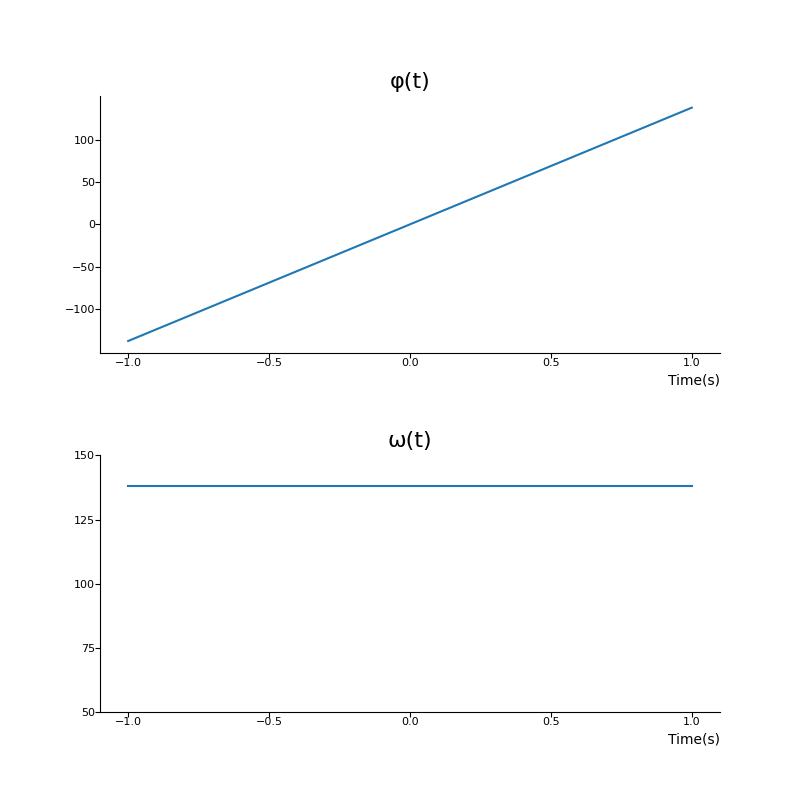}}
    \caption{Example 3, $\omega(t)$ is located at the geometric center of the Fourier spectrum}\label{fig:fig_4} 

\end{figure}


In Example 3, Figure \ref{fig:fig_4}(a) shows that the range of the Fourier spectrum of $\Psi_{u}(t)$ is 0.5Hz$\sim$43.5Hz. On the other hand, Figure \ref{fig:fig_4}(c) shows that $\Psi_{u}(t)$ has a constant intrinsic instantaneous frequency of 22Hz (138.16rad/s), which is located at the geometric center of the Fourier spectrum. Figure\ref{fig:fig_4}(b) shows that $\Psi_{u}(t)$ is the signal with energy concentration in the time domain. It can be proven that $\Psi_{u}(t)$ in Example 3 satisfies the definition of the wavelet function.

In the above examples, the intrinsic instantaneous frequency is calculated using the parity decomposition of the signals. If the traditional Hilbert transform analytical function method were used for the calculation, the results would be the same. However, the calculation of intrinsic phase and intrinsic instantaneous frequency based on the parity decomposition of the signals is relatively small.

"Keeping the monotonicity of its intrinsic phase function throughout the entire time range" can be used as the criterion to determine the maximum width of the frequency band occupied by a certain mode. However, finding the center frequency of modes is still a problem. Using the real part spectrum $Fre(\omega)$ and the imaginary part spectrum $Fim(\omega)$ mentioned above, the center frequency point of the mode can be located.

\section{Searching for the Modes}
\label{sec:search_mode}

A mode is a narrow band function, and if the center frequency and bandwidth of the mode are found, the question raised at the end of the section\ref{subsection:mode_extraction} of this article is answered. The center frequency of the mode can be found using the imaginary part spectrum and the real part spectrum of the signal, and the bandwidth of the mode can be determined by calculating the intrinsic phase function and the instantaneous frequency of the signal.

\subsection{The role of the imaginary part spectrum and the real part spectrum}

Comparing the real part spectrum and the imaginary part spectrum of finite-length real signals will help find the center frequency of the modes.
The following examples reveal some properties of the real imaginary part spectrum of the finite-length discrete signals. 

\vspace{1mm}

\textbf{Example 4:}$\Psi_{u}(t)=1.3sin(5\pi t)+sin(12\pi t)+cos(4\pi t)+cos(13\pi t), t \in [-1,1]$.

We sample $\Psi_{u}(t)$ at 100Hz.

\begin{figure}[ht]
\centering
\includegraphics[width=7cm,height=3.2cm]{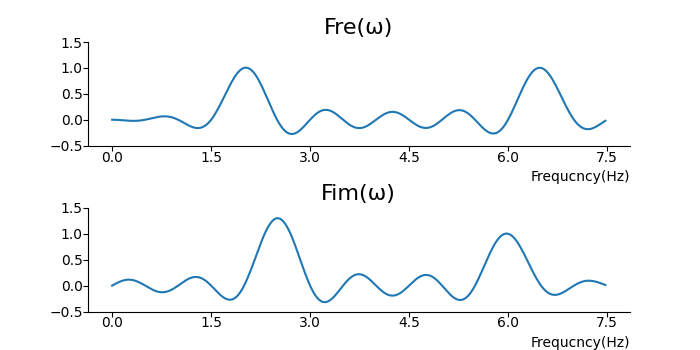}

\caption{\label{fig:fig_5} Example 4, $Fre(\omega)$ and $Fim(\omega)$ each has two principal lobes}
\end{figure}

In Example 4, the real part spectrum $Fre(\omega)$ and the imaginary part spectrum $Fim(\omega)$ of $\Psi_{u}(t)$ are shown in Figure \ref{fig:fig_5}. The two principal lobes of $Fre(\omega)$ correspond to the single-frequency components $cos(4\pi t)$ and $cos(13\pi t)$ respectively.
The two principal lobes of $Fim(\omega)$ correspond to the single-frequency components $sin(5\pi t)$ and $sin(12\pi t)$ respectively.
The duration of $\Psi_{u}(t)$ is 2 seconds, so the minimum frequency resolution unit $\epsilon$ is 0.5Hz. In Figure \ref{fig:fig_5}. The width of the support intervals of the principal lobes of $Fre(\omega)$ and $Fim(\omega)$ are 1Hz, twice the minimum frequency resolution unit $\epsilon$. 
The width of the positive and negative half of the side lobes is 0.5Hz, the same as the frequency domain resolution unit $\epsilon$.
It can be proven that when the individual harmonic components of the original signal are far apart in the frequency domain, the above conclusions about the width of the principal and side lobes of the spectrum are always correct.

\vspace{1mm}

\textbf{Example 5:}$\Psi_{u}(t)=1.3sin(5\pi t)+sin(7\pi t)+cos(4\pi t)+cos(6\pi t), t \in [-1,1]$.

We sample $\Psi_{u}(t)$ at 100Hz.

\begin{figure}[ht] 
    \centering 
    \subfloat[]{        \includegraphics[width=3.1cm,height=3.5cm]{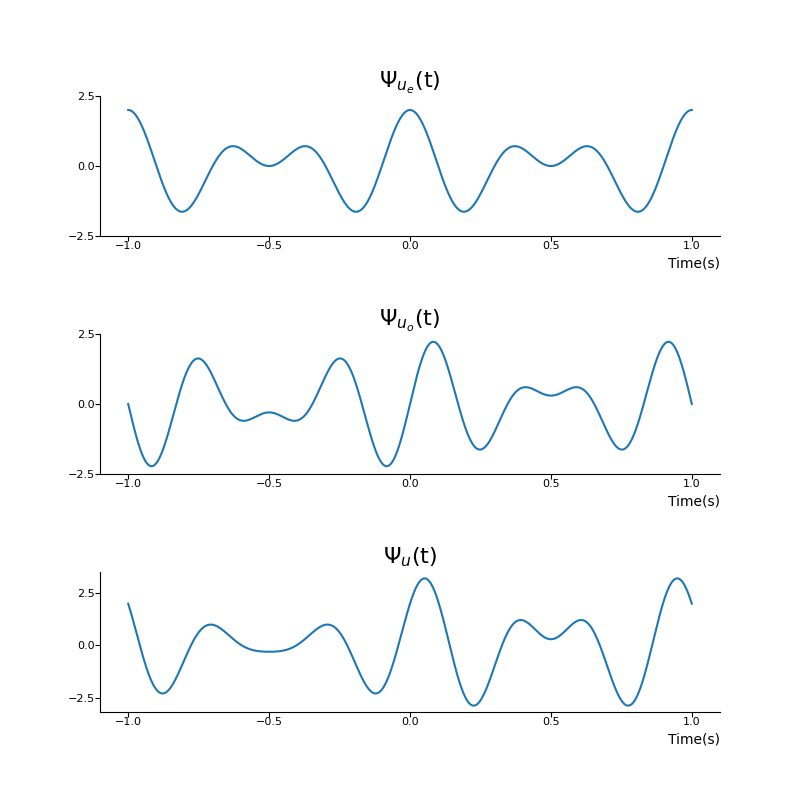}}%
    \hspace*{-3mm} 
    \subfloat[]{        \includegraphics[width=3.1cm,height=3.5cm]{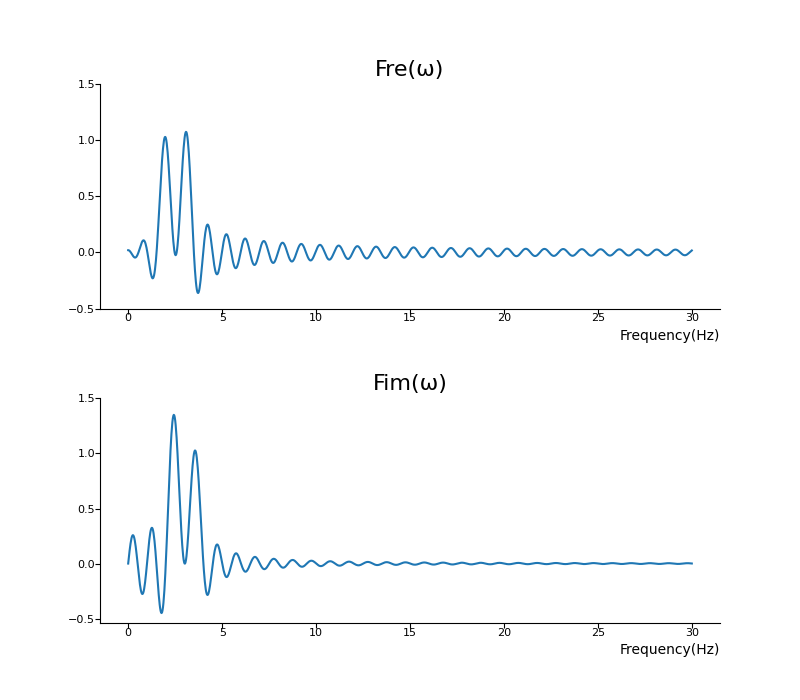}}     
    \hspace*{-3mm} 
    \subfloat[]{        \includegraphics[width=3.1cm,height=3.5cm]{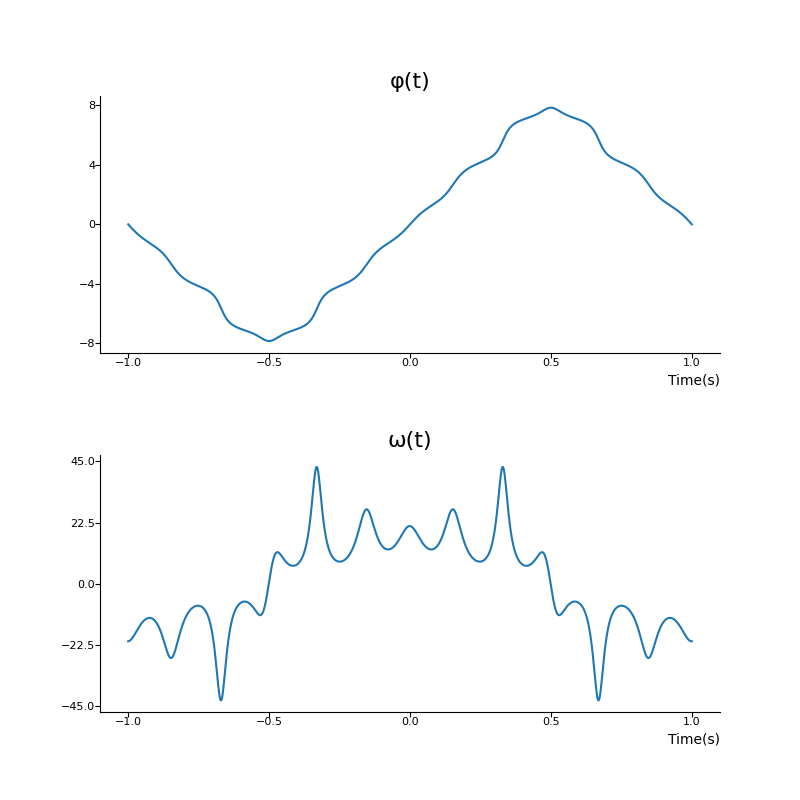}}%
    \caption{Example 5, The phase function $\varphi(t)$ is not monotonic}\label{fig:fig_6} 
\end{figure}

In Example 5, the real part spectrum $Fre(\omega)$ and imaginary part spectrum $Fim(\omega)$ are shown in \ref{fig:fig_6} (b). The harmonic components specified in the real part spectrum $Fre(\omega)$ are $cos(4\pi t)$ and $cos(6\pi t)$ with a frequency difference of 1Hz. This frequency difference is twice the minimum resolution unit $\epsilon$. The harmonic components specified in the imaginary part spectrum $Fim(\omega)$ are $1.3sin(5\pi t)$ and $sin(7\pi t)$ and the frequency difference is 1Hz. This is also twice the minimum frequency resolution unit $\epsilon$. The principal lobes corresponding to each single-frequency component can appear completely and independently in the real part spectrum and the imaginary part spectrum, as shown in \ref{fig:fig_6} (b). Figure \ref{fig:fig_6} (c) shows that the phase function $\varphi(t)$ is not monotonic, which is caused by the fact that the various harmonic components contained in $\Psi_{u}(t)$ are slightly distant from each other in the frequency domain. In this example, all single-frequency components in $\Psi_{u}(t)$ are independent modes. This is different from Examples 2 and 3 in this article. In these examples, several single-frequency components form a common principal lobe in the frequency domain, the mode in the time domain.

If a principal lobe wider than 2 times the frequency resolution unit $\epsilon$ exists in the real part spectrum and the imaginary part spectrum of the signal, there must be multiple harmonic components in it. Each mode corresponds to one or more adjacent principal lobes in the real imaginary part spectrum, and the center frequency of the mode can be determined according to the position of the principal lobe in the spectrum.

\subsection{Finding the boundary of the frequency band of the mode}

An isolated principal lobe in the spectrum corresponds to an independent mode.

If there are several overlapping principal lobes, we should take the support interval width of one of the principal lobes as the initial value of the mode frequency bandwidth. In this case, an iterative calculation is needed to calculate the actual bandwidth occupied by the modes. In the process of iteration, new frequency components are constantly added. Each iteration needs to calculate its phase function in the time domain. As long as the phase function remains monotonous, the iteration continues. Until it is found that the intrinsic phase function is no longer monotonic, the iterative operation stops. At this point, the sum of all frequency components belonging to this mode is realized to the greatest extent, and the mode is found. From a view of the frequency domain, this operation is to find the frequency interval corresponding to the mode. The specific calculation steps are as follows:

\begin{enumerate}
  \item Based on the real part spectrum $Fre(\omega)$ and the imaginary part spectrum $Fim(\omega)$ of the signal $\Psi_{u}(t)$, the frequency band $[\Omega_{c1},\Omega_{c2}]$ of a certain principal lobe can be determined.
 
  \item In the frequency band ($[\Omega_{c1},\Omega_{c2}]$) using the orthogonal projection operator given in equation (\ref{eq:mode_local} )of this article, the component $\Psi_u^{[\Omega_{c1}, \Omega_{c2}]}(t)$ of the signal $\Psi_{u}(t)$  can be calculated.
  
  \item The intrinsic phase function $\varphi^{[\Omega_{c1},\Omega_{c2}]}(t)$ and the intrinsic instantaneous frequency $\omega^{[\Omega_{c1}, \Omega_{c2}]}(t)$ of $\Psi_u^{[\Omega_{c1}, \Omega_{c2}]}$ can be calculated .

   \item The lower limit of the frequency band is changed from $\Omega_{c1}$ to $\Omega_{c1}-\epsilon$. In the extended frequency band $[\Omega_{c1}-\epsilon,\Omega_{c2}]$, we can repeat steps 2 and 3 to calculate the intrinsic phase function $\varphi^{[\Omega_{c1}-\epsilon,\Omega_{c2}]}(t)$ and the intrinsic instantaneous frequency $\omega^{[\Omega_{c1}-\epsilon,\Omega_{c2}]}(t)$. $\epsilon$ is the smallest resolution unit in the frequency domain.

  \item The upper limit of the frequency band is changed from $\Omega_{c2}$ to $\Omega_{c2}+\epsilon$. In the extended frequency band $[\Omega_{c1},\Omega_{c2}+\epsilon]$, we can repeat steps 2 and 3 to calculate the intrinsic phase function $\varphi^{[\Omega_{c1},\Omega_{c2}+\epsilon]}(t)$ and the intrinsic instantaneous frequency $\omega^{[\Omega_{c1},\Omega_{c2}+\epsilon]}(t)$. $\epsilon$ is the smallest resolution unit in the frequency domain.
\end{enumerate}

The above process is iteratively executed until the following situation occurs:

If there is always $\omega^{[\Omega_{c1}, \Omega_{c2}]}(t)>0$ in the entire time segment and neither $\omega^{[\Omega_{c1}-\epsilon,\Omega_{c2}]}(t)$ nor $\omega^{[\Omega_{c1},\Omega_{c2}+\epsilon]}(t)$ meets this condition, then $\Psi_u^{[\Omega_{c1}, \Omega_{c2}]}(t)$ is the intrinsic positive frequency mode in the band $[\Omega_{c1}, \Omega_{c2}]$.
If there is always $\omega^{[\Omega_{c1}, \Omega_{c2}]}(t)<0$ in the entire time segment and neither $\omega^{[\Omega_{c1}-\epsilon,\Omega_{c2}]}(t)$ nor $\omega^{[\Omega_{c1},\Omega_{c2}+\epsilon]}(t)$ meets this condition, then $\Psi_u^{[\Omega_{c1}, \Omega_{c2}]}(t)$ is the intrinsic negative frequency mode in the band $[\Omega_{c1}, \Omega_{c2}]$;

From the above calculation process, we know that the intrinsic mode is extracted locally in the frequency domain, allowing any intrinsic mode of interest to be extracted directly. Moreover, searching for mode position and extracting mode are the same process, which greatly reduces the complexity of the calculation.

\subsection{The method of extracting modes on imaginary part and real part respectively }

The algorithm introduced in this paper first performs parity decomposition for the original finite length signal in the time domain, the real part spectrum and the imaginary part spectrum can be obtained separately. The real part spectrum is the spectrum of the even component of the original signal. The imaginary part spectrum is the spectrum of the odd component of the original signal. This indicates that extracting the modes in the real and imaginary part spectrum is feasible and effective. The modes obtained in this way are named real part mode and imaginary part mode of the signal. The procedures of decomposing real part modes and imaginary part modes are similar to those described above. The method of calculating intrinsic phase and instantaneous frequency for the pure real part modes or the pure imaginary part modes is already detailed in Section\ref{sec:inherent_frequency}. The mode decomposition result obtained using this method is also orthogonal and unique in the mode set.

\subsection{The mode decomposition examples}

\vspace{1mm}

\textbf{Example6:}  $\Psi_{u}(t)=0.6cos(5\pi t)+1.5cos(6\pi t)+1.3cos(28\pi t)+sin(7\pi t)+1.2sin(22\pi t)+a(t)*1.5sin(62\pi t)$,
$t \in [-1,1]$.

Among them, $a(t)$ is a window function. when $t \in [-0.5,0.5]$ $a(t)=1$, else $a(t)=0$ .

We sample $\Psi_{u}(t)$ at 100Hz. 
\begin{figure}[ht] 
    \centering 
    \subfloat[]{        \includegraphics[width=4.2cm,height=3.5cm]{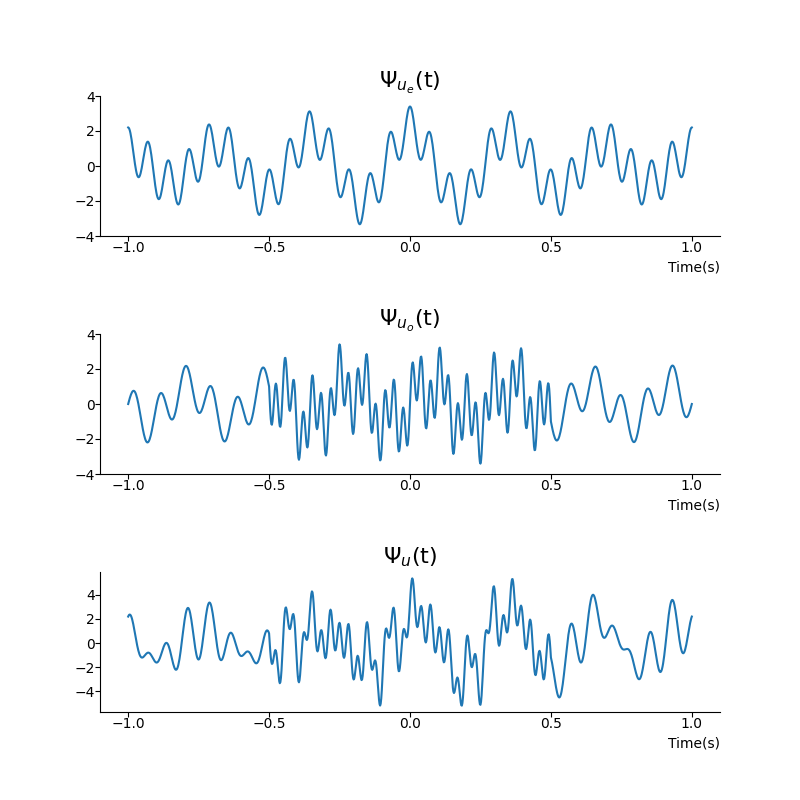}}  
    \hspace*{-3mm} 
    \subfloat[]{        \includegraphics[width=4.2cm,height=3.5cm]{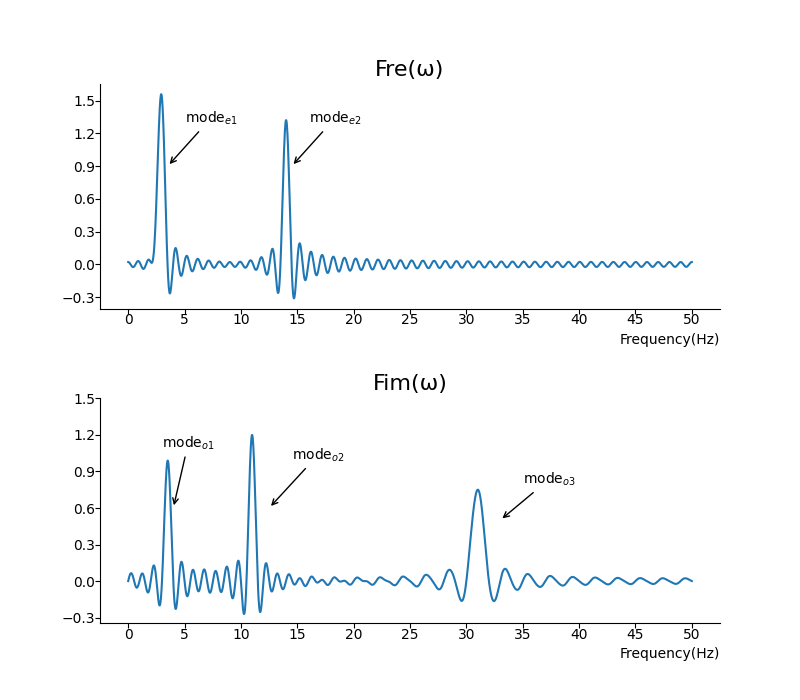}}%
    
    \caption{Example 6, The even and odd component and spectrum of $\Psi_{u}(t)$}\label{fig:fig_7} 
\end{figure}

Figure \ref{fig:fig_7}(a) shows $\Psi_{u}(t)$ as well as its even component $\Psi_{u_e}(t)$ and its odd component $\Psi_{u_o}(t)$. Figure\ref{fig:fig_7}(b) shows the real part spectrum $Fre(\omega)$ and the imaginary part spectrum $Fim(\omega)$. The real part spectrum has two principal lobes, corresponding to two modes $mode_{e1}$ and $mode_{e2}$. $mode_{e1}$ contains two single-frequency components $0.6cos(5\pi t)$ and $1.5cos(6\pi t)$. These two single-frequency components are so close together in the frequency domain that they form a mode. $mode_{e2}$ contains only one single-frequency component $1.3cos(28\pi t)$. The frequency bands [2Hz, 3Hz] and [13Hz, 15Hz] are selected to calculate $mode_{e1}$ and $mode_{e2}$ respectively. Figure\ref{fig:fig_7}(b) also shows that the imaginary part spectrum $Fim(\omega)$ has three main lobes, corresponding to three modes $mode_{o1}$ and $mode_{o2}$ and $mode_{o3}$, respectively. $mode_{o1}$ is the single-frequency component $sin(7\pi t)$. $mode_{o2}$ is the single-frequency component $1.2sin(22\pi t)$. $mode_{o3}$ is the short-term component $a (t)*1.5sin(62\pi t)$. Because $mode\_{o3}$ is a short-term component in the time domain, its principal lobe is wider than the other two modes in the frequency domain. The bands [2.5Hz,3.5Hz] and [10Hz,12Hz] and [26Hz,36Hz] are selected to calculate $mode_{o1}$ and $mode_{o2}$ and $mode_{o3}$ respectively. 

\begin{figure}[ht] 
    \centering 
    \subfloat[]{        \includegraphics[width=4.2cm,height=3.5cm]{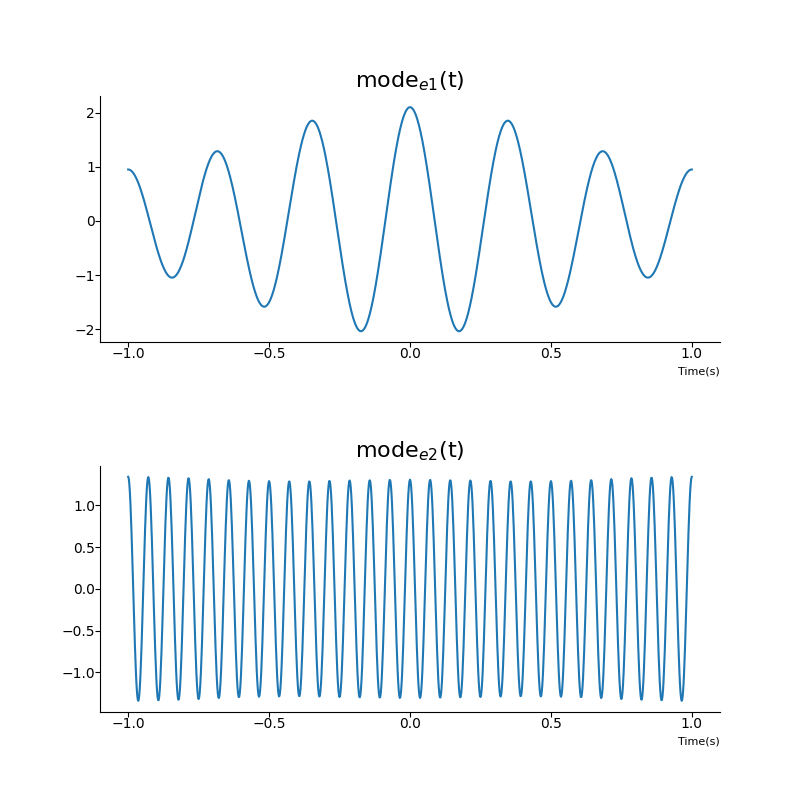}}%
    \hspace{-3mm} 
    \subfloat[]{        \includegraphics[width=4.2cm,height=3.5cm]{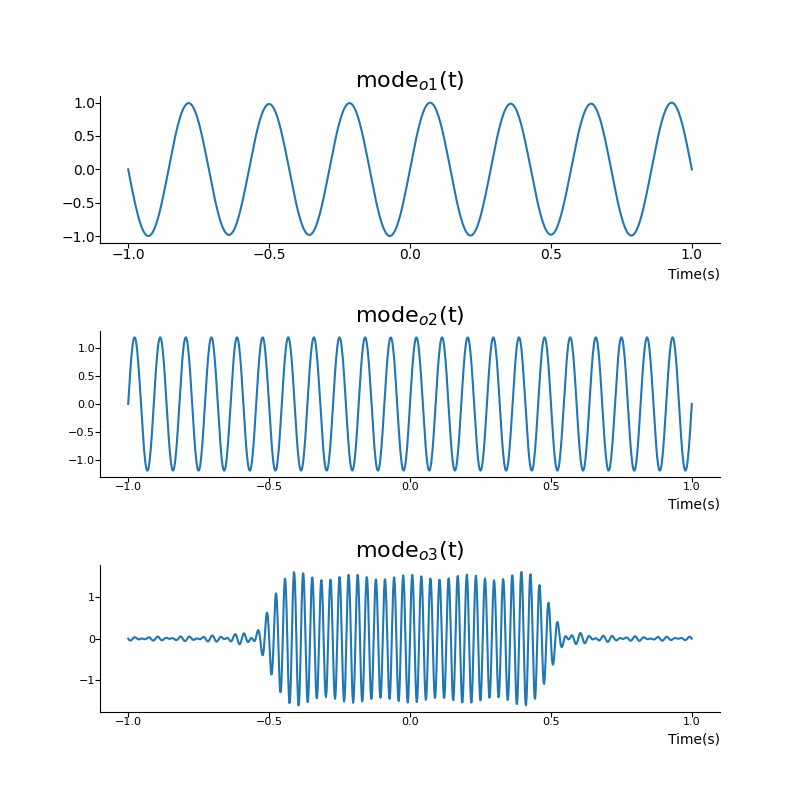}}      
    
    \caption{Example 6,$\Psi_{u}(t)$'s real part modes and imaginary part modes  }\label{fig:fig_8} 
\end{figure}

The calculation results for all modes are shown in \ref{fig:fig_8}.

Figure \ref{fig:fig_8} shows that the real part mode $mode_{e1}$ is an amplitude modulation wave, the imaginary part mode $mode_{o3}$ is a short-term wave and all other modes are single-frequency signals. The calculated results are consistent with the analysis. The orthogonal mode decomposition method is also local in the time domain. The short-term signal in the time domain usually becomes an intrinsic mode, just like $mode_{o3}$.

\section{Extraction of the non-oscillatory low-frequency components }
\label{sec:low_freq}

Whether according to the traditional definition of the mode or the definition of intrinsic mode given in this paper, a mode is a signal with an oscillating shape. However, actual physical signals generally contain non-oscillating low-frequency components.

The previous mode decomposition algorithms first extract the oscillating modes one by one leaving the non-oscillating low-frequency component at the end, because these mode decomposition methods are global algorithms. However, the orthogonal mode decomposition method proposed in this paper is a local mode extraction method, which can extract any interested mode without calculating other modes. The non-oscillating low-frequency components of the original signal can also be extracted directly, and the calculation amount can be reduced as much as possible.

 The non-oscillating low-frequency component $imf\_0$ is in the frequency band $[0,\Omega_{c1}]$. That is, the upper limit of the bandwidth is $\Omega_{c1}$. A new sampling period $\Delta_0$ can be selected for resampling the interpolation function $\Psi_{u}(t)$.
The interpolation function $\Psi_{u}(t)$ of the original finite-length discrete signal $\boldsymbol{u}$ is resampled to form a new sequence with fewer members. The resampling period $\Delta_0$ is subject to the following constraints:
      $$\Delta < \Delta_0 < \frac{\pi}{\Omega_{c1}}$$      

$\Delta$ is the sampling period of the original discrete signal $\boldsymbol{u}$. The resampling period $\Delta_0$ should be greater than $\Delta$, but less than the upper limit value $\frac{\pi}{\Omega_{c1}}$ determined by the low-frequency bandwidth. According to the sampling theorem, the resampled sequence retains all the information features of the non-oscillatory low-frequency component $imf_0$. The resampled sequence is:
$$\boldsymbol{u}_0=[u_0(-l_0),u_0(-l_0+1),\hdots,u_0(0),\hdots,u_0(l_0-1),u_0(l_0)]$$

The members of $\boldsymbol{u}_0$ are obtained by resampling the original interpolation function $\Psi_{u}(t)$, and the length of $\boldsymbol{u}_0$ is $n_0=2l_0+1$. The resampled sequence $\boldsymbol{u}_0$ can also be decomposed into even subsequence $\boldsymbol{u}_{0e}$ and odd subsequence $\boldsymbol{u}_{0o}$ in the way of equation (\ref{orth_decomp}) in this article. 

The polynomial $\Psi_{u_{0e}}(t)$ can be used to fit the even subsequence $\boldsymbol{u}_{0e}$, and the polynomial $\Psi_{u_{0o}}(t)$ can be used to fit the odd subsequence $\boldsymbol{u}_{0o}$. It can be assumed that:
\begin{align}
   \begin{split}    
    \Psi_{u_{0e}}(t) &= a_0+a_1t^2+a_2t^4+\hdots+a_{l_0}t^{2l_0} \\
    \Psi_{u_{0o}}(t) &= b_1t+b_2t^3+b_3t^5\hdots+b_{l_0}t^{2l_0-1}\nonumber
   \end{split}
  \quad \quad \quad \text{$t \in [-1,1]$} 
\end{align}

Among them, $\{a_i | i=0,1,\hdots,l_0\}$ and $\{b_i | i=1,\hdots,l_0\}$ are the real coefficients. $\Psi_{u_{0e}}(t)$ is symmetric and  $\Psi_{u_{0o}}(t)$ is anti-symmetric, so only half of the entire segment needs to be analyzed. $t\in [0,1]$.

Obviously, $\Psi_{u_{0e}}(0)=a_0$. The other coefficients $\{a_i | i=1,\hdots,l_0\}$ can be obtained from the linear equation system (\ref{Matrix_multiplication_e}):
\begin{equation} \label{Matrix_multiplication_e}
    \begin{bmatrix}
        {\Delta_0}^2 &  {\Delta_0}^4  & \hdots & {\Delta_0}^{2l_0} \\
        2^2{\Delta_0}^2 & 2^4{\Delta_0}^4  & \hdots & 2^{2l_0}{\Delta_0}^{2l_0} \\
        \vdots & \vdots & \ddots & \vdots\\
        {l_0}^2{\Delta_0}^2 & {l_0}^4{\Delta_0}^4 & \hdots & {l_0}^{2l_0}{\Delta_0}^{2l_0}
    \end{bmatrix}
    \begin{bmatrix}
        a_1  \\
        a_2 \\
        \vdots \\
        a_{l_0}
    \end{bmatrix}    
     =
    \begin{bmatrix}
       \overset{\sim}{\Psi}_{u_{0e}}(\Delta_0) \\
       \overset{\sim}{\Psi}_{u_{0e}}(2\Delta_0) \\
       \vdots\\
       \overset{\sim}{\Psi}_{u_{0e}}(l_0\Delta_0) 
    \end{bmatrix}
\end{equation}

Among them:

$$\overset{\sim}{\Psi}_{u_{0e}}(i\Delta_0)={\Psi}_{u_{0e}}(i\Delta_0)-\Psi_{u_{0e}}(0) \hspace{3mm}|i=1,\hdots,l_0$$

The coefficients $\{b_i | i=1,\hdots,l_0\}$ can be obtained from another linear equation system\ref{Matrix_multiplication_o}):

\begin{equation} \label{Matrix_multiplication_o}
    \begin{bmatrix}
        {\Delta_0} &  {\Delta_0}^3  & \hdots & {\Delta_0}^{2l_0-1} \\
        2{\Delta_0} & 2^3{\Delta_0}^3  & \hdots & 2^{2l_0-1}{\Delta_0}^{2l_0-1} \\
        \vdots & \vdots & \ddots & \vdots\\
        {l_0}{\Delta_0} & {l_0}^3{\Delta_0}^3 & \hdots & {l_0}^{2l_0-1}{\Delta_0}^{2l_0-1}
    \end{bmatrix}
    \begin{bmatrix}
        b_1  \\
        b_2 \\
        \vdots \\
        b_{l_0}
    \end{bmatrix}    
     =
    \begin{bmatrix}
       \Psi_{u_{0o}}(\Delta_0) \\
       \Psi_{u_{0o}}(2\Delta_0) \\
       \vdots\\
       \Psi_{u_{0o}}(l_0\Delta_0) 
    \end{bmatrix}
\end{equation}

\smallskip

The $l_0$-order coefficient matrices of linear equation systems(\ref {Matrix_multiplication_e})and(\ref{Matrix_multiplication_o}) are full rank.

Since the resampling period $\Delta_0$ only depends on the non-oscillating low-frequency component's bandwidth $\Omega_{c1}$, $\Delta_0$ is much larger than the original discrete signal's sampling period $\Delta$, so the linear equation systems' order $l_0$ is much smaller than half of the length of the original discrete signal. The above resampling method reduces the complexity of calculating non-oscillatory low-frequency components.

\section{The comparison of several mode decomposition methods}
\label{sec:comparison}

The orthogonal mode decomposition method given in this paper is essentially a local mode extraction algorithm, which determines that this algorithm has the characteristics of small computation. The results of mode decomposition are unique and orthogonal.
Some examples from the reference\cite{dragomiretskiy2013variational} are cited to compare the orthogonal mode decomposition method with the empirical mode decomposition (EMD) and the variational mode decomposition (VMD).

\vspace{1mm}

\textbf{Example7:}  $\Psi_{u}(t)=6t+cos(8\pi t)+0.5cos(40\pi t) \quad t\in [-1,1]$.

We sample $\Psi_{u}(t)$ at 100Hz.

\begin{figure}[ht] 
    \centering 
    \subfloat[]{        \includegraphics[width=4.2cm,height=3.6cm]{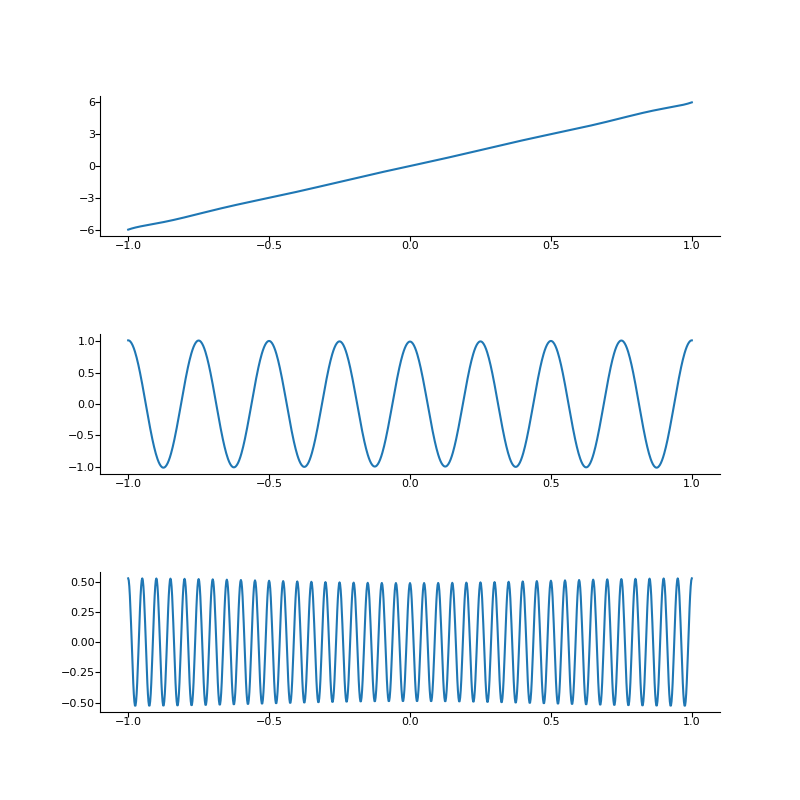}}%
    \hspace*{-4mm} 
    \subfloat[]{        \includegraphics[width=2.5cm,height=3.6cm]{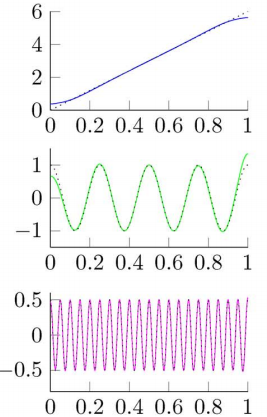}}%
    \hspace*{-3mm} 
    \subfloat[]{        \includegraphics[width=2.5cm,height=3.6cm]{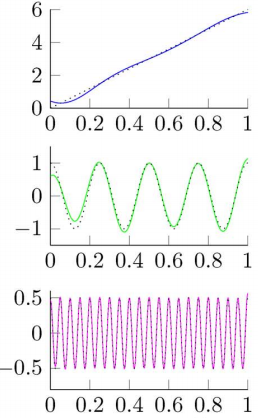}}%
    \caption{Example 7, three methods are used for the mode decomposition of $\Psi_{u}(t)$}\label{fig:fig_9} 
\end{figure}

Example 7 is cited from reference \cite{dragomiretskiy2013variational}, which presented the results of using VMD and EMD for $\Psi_{u}(t)$ decomposition. The orthogonal mode decomposition is used in this article. Figure \ref{fig:fig_9} shows the results obtained using the above three methods. Figure \ref{fig:fig_9}(a) shows the results of using the orthogonal mode decomposition, Figure\ref{fig:fig_9}(b) shows the results of using the variational mode decomposition (VMD), and Figure \ref{fig:fig_9}(c) shows the results of using the empirical mode decomposition (EMD). Figures\ref{fig:fig_9}(b) and \ref{fig:fig_9}(c) are cited from references\cite{dragomiretskiy2013variational}.

\textbf{Example8:}  $\Psi_{u}(t)=6t^2+\cos(10\pi|t|+ 10\pi  t^2)+\psi(t)$

Among them:

$\psi(t)=
\begin{cases}
       cos(60\pi t)  &\text{ $-0.5<=t<=0.5$ }\\
      cos(80\pi t-10\pi)  &\text{$-1<t<-0.5$ or $0.5<t<1$}
\end{cases}$ 

\vspace{2mm} 

 We sample $\Psi_{u}(t)$ at 100Hz.
\begin{figure}[ht] 
    \centering 
    \subfloat[]{        \includegraphics[width=4.2cm,height=3.6cm]{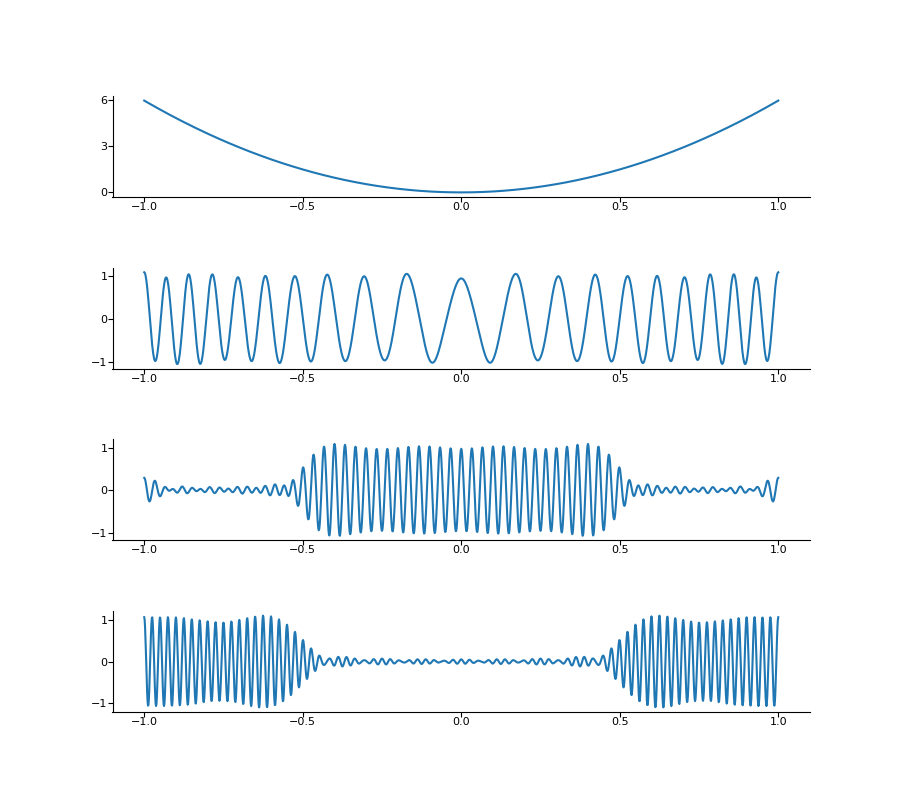}}%
    \hspace*{-4mm} 
    \subfloat[]{        \includegraphics[width=2.5cm,height=3.6cm]{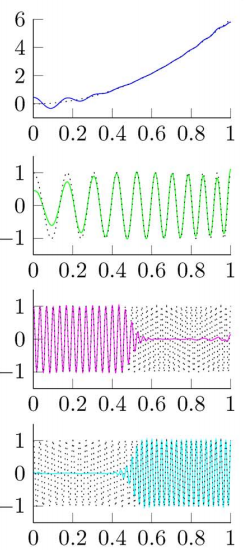}}%
    \hspace*{-3mm}
    \subfloat[]{        \includegraphics[width=2.5cm,height=3.6cm]{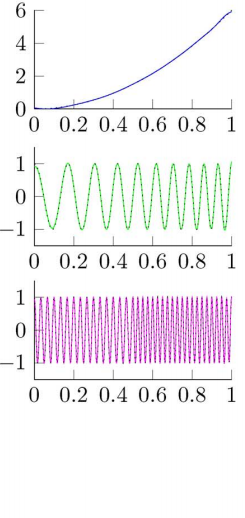}}%
    \caption{Example 8, three methods are used for the mode decomposition of $\Psi_{u}(t)$}\label{fig:fig_10} 

\end{figure}

Example8 is cited from reference \cite{dragomiretskiy2013variational}, which presented the results of using VMD and EMD for $\Psi_{u}(t)$ decomposition. The orthogonal mode decomposition is used in this article. Figure \ref{fig:fig_10} shows the results obtained using the above three methods. Figure \ref{fig:fig_10}(a) shows the results of using the orthogonal mode decomposition, Figure \ref{fig:fig_10}(b) shows the results of using the variational mode decomposition (VMD), and Figure \ref{fig:fig_10}(c) shows the results of using the empirical mode decomposition (EMD). Figures\ref{fig:fig_10}(b) and \ref{fig:fig_10}(c) are cited from references\cite{dragomiretskiy2013variational}.

 \medskip
 
\textbf{Example9:}  $\Psi_{u}(t)=a_{u}(t)+b_{u}(t)\quad   t\in [-1,1]$  

Among them:
\begin{align}
   \begin{split}    
    a_u(t)&=\frac{1}{1.2+cos(2\pi t)}\nonumber\\
    b_u(t)&=\frac{cos(32\pi |t|+0.2cos(64\pi t))}{1.5+sin(2\pi | t |)}\nonumber
   \end{split}  
\end{align}

 We sample $\Psi_{u}(t)$ at 100Hz.

\begin{figure}[ht] 
    \centering 
    \subfloat[]{        \includegraphics[width=4.2cm,height=3.4cm]{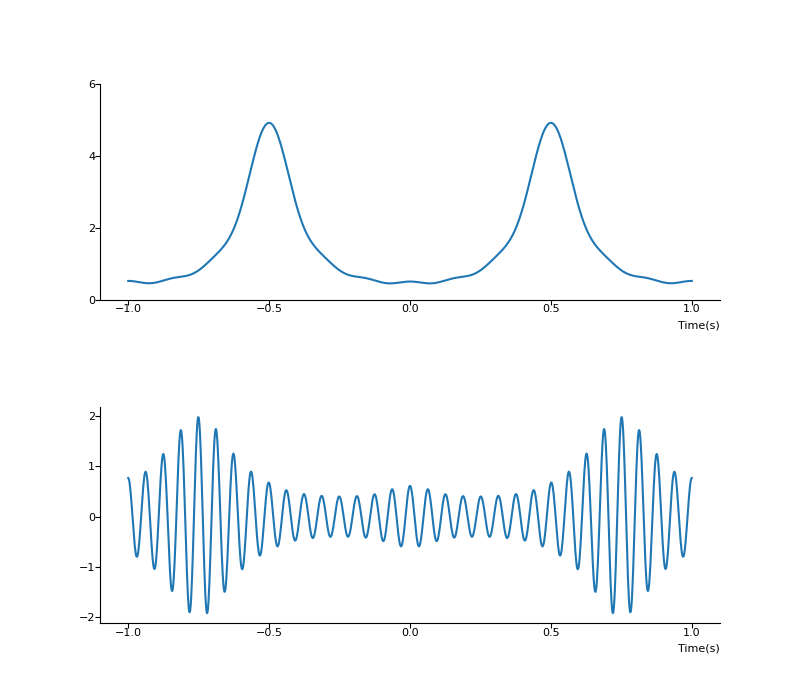}}%
    \hspace*{-4mm} 
    \subfloat[]{        \includegraphics[width=2.5cm,height=3.4cm]{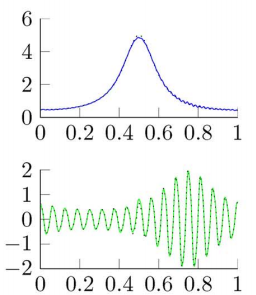}}%
    \hspace*{-3mm} 
    \subfloat[]{        \includegraphics[width=2.5cm,height=3.4cm]{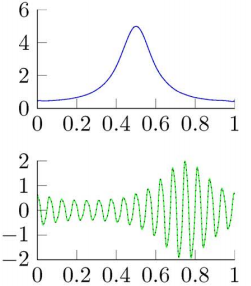}}%
    \caption{Three methods are used for the mode decomposition of $\Psi_{u}(t)$ in Example 9}\label{fig:fig_11} 

\end{figure}

Example9 is cited from reference \cite{dragomiretskiy2013variational}, which presented the results of using VMD and EMD for $\Psi_{u}(t)$ decomposition. The orthogonal mode decomposition is used in this article. Figure\ref{fig:fig_11} shows the results obtained using the above three methods. Figure \ref{fig:fig_11}(a) shows the results of using the orthogonal mode decomposition, Figure\ref{fig:fig_11}(b) shows the results of using the variational mode decomposition (VMD), and Figure\ref{fig:fig_11}(c) shows the results of using the empirical mode decomposition (EMD). Figures\ref{fig:fig_11}(b) and \ref{fig:fig_11}(c) are cited from references\cite{dragomiretskiy2013variational}.

\textbf{Example10:} $\Psi_{u}(t)=f_{saw}(t)+0.3sin(72\pi t) \quad \quad   t\in [-1,1]$ , $\Psi_{u}(t)$ is sampled at a frequency of 276Hz.

In Example10, $\Psi_{u}(t)$ contains the sawtooth wave $f_{saw}(t)$ and the single-frequency component $0.3sin(72\pi t)$. $\Psi_{u}(t)$ and its all components are odd functions. A similar example is provided in reference \cite{dragomiretskiy2013variational}, which shows the waveforms of the modes obtained through EMD and VMD, respectively. This article uses the orthogonal mode decomposition algorithm to perform the mode decomposition of $\Psi_{u}(t)$ in Example10.

\begin{figure}[ht] 
    \centering
    \subfloat[]{        \includegraphics[width=4.5cm,height=3.4cm]{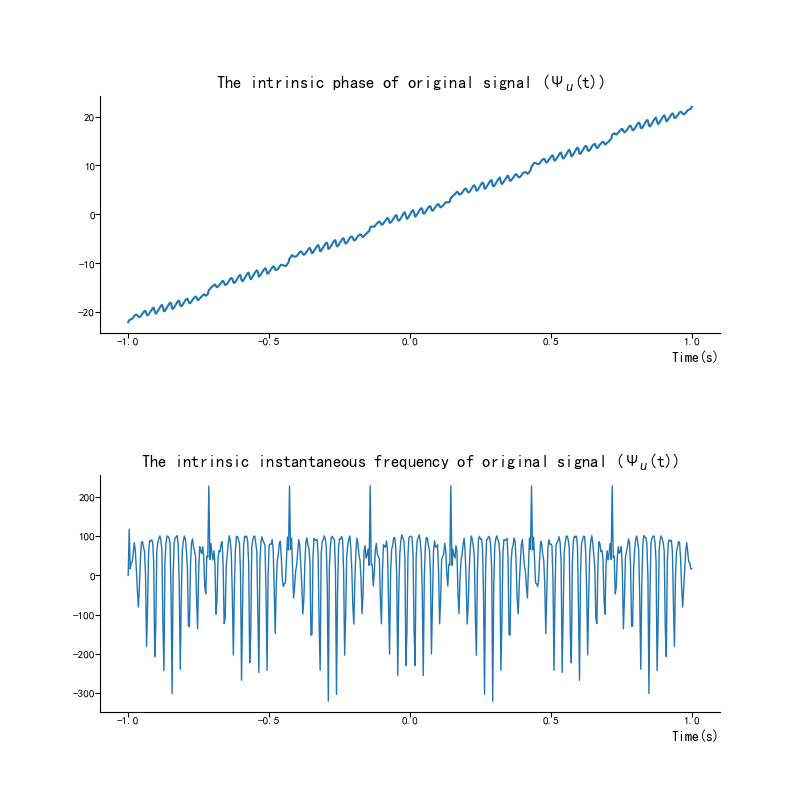}}%
    \hspace*{-3mm}
    \subfloat[]{        \includegraphics[width=4.5cm,height=3.4cm]{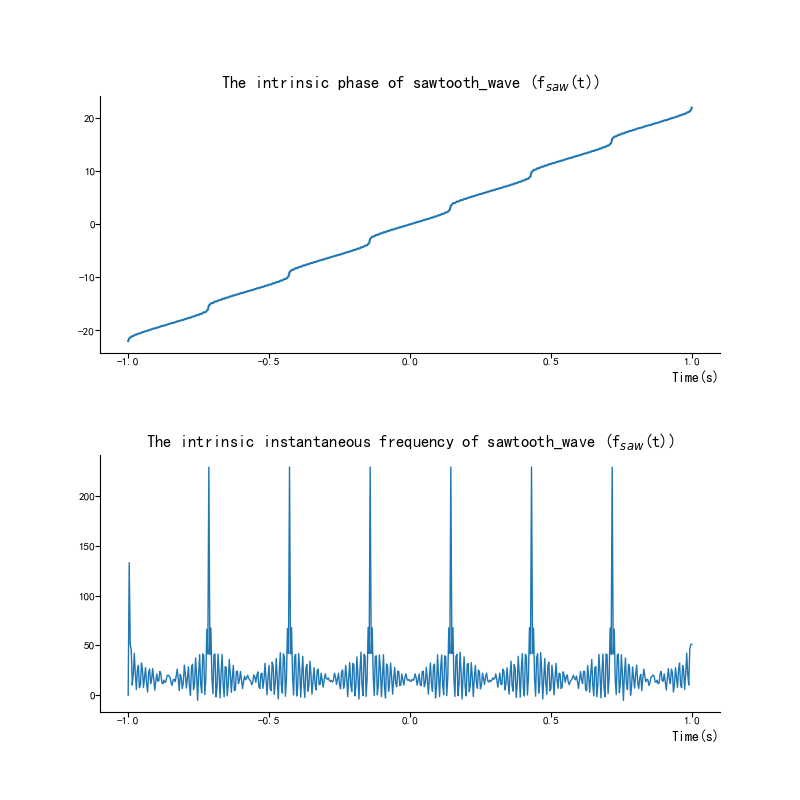}}%
    \caption{In Example 10, $f_{saw}(t)$ is a mode , while $\Psi_{u}(t)$ is not }\label{fig:fig_12} 
\end{figure}

Figure\ref{fig:fig_12}(a) shows the intrinsic phase and the instantaneous frequency of $\Psi_{u}(t)$ in Example 10. Since $\Psi_{u}(t)$ contains a sawtooth wave component $f_{saw}(t)$ and a single-frequency component $0.3sin(72\pi t)$, its intrinsic phase is not monotonic and its instantaneous frequency cannot remain positive throughout the time segment. Figure \ref{fig:fig_12}(b) is the intrinsic phase function and the instantaneous frequency of the sawtooth component $f_{saw}(t)$. The intrinsic phase function of $f_{saw}(t)$ has monotonicity, and the instantaneous frequency of $f_{saw}(t)$ remains positive throughout the time segment. The sawtooth $f_{saw}(t)$ is one mode, and the single-frequency component $0.3sin(72\pi t)$ is another mode. $\Psi_{u}(t)$ in example 10 contains the two modes.


\begin{figure}[ht] 
    \centering 
    \subfloat[]{        \includegraphics[width=4.5cm,height=3.6cm]{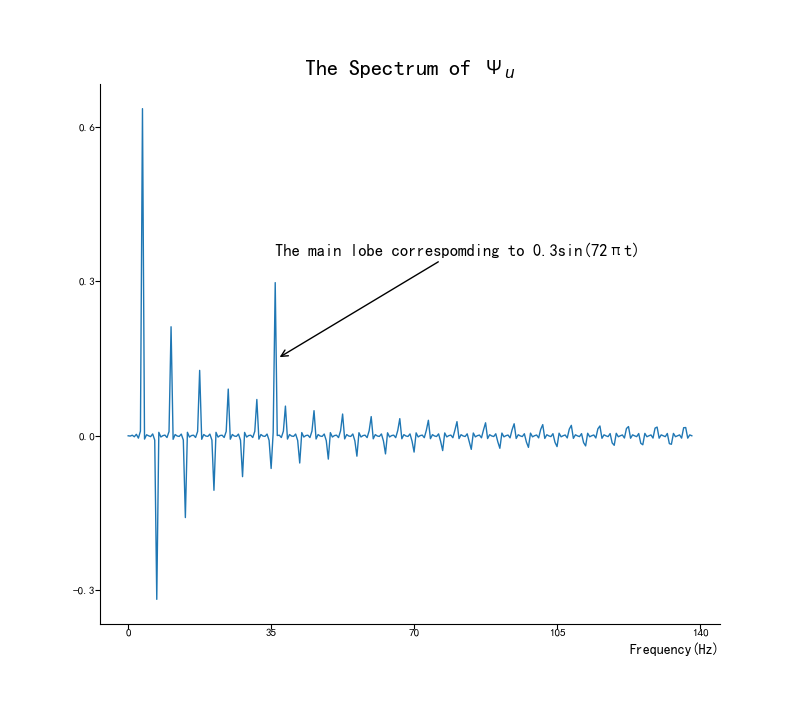}}%
    \hspace*{-3mm} 
    \subfloat[]{        \includegraphics[width=4.5cm,height=3.6cm]{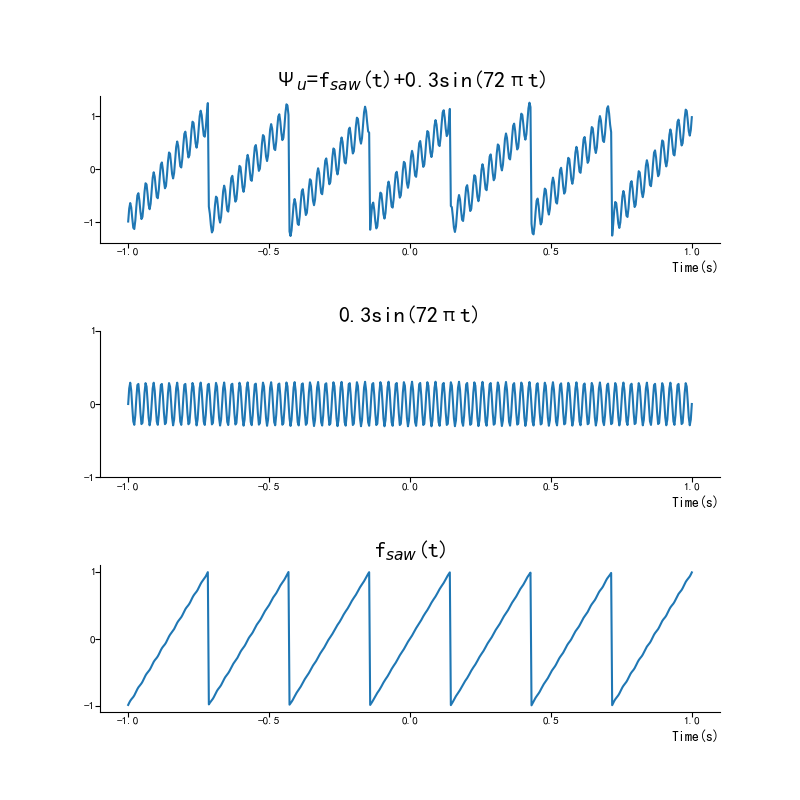}}%
    \caption{$0.3sin(72\pi t)$ is found in the spectrum, and $f_{saw}(t)$ is restored }\label{fig:fig_13} 
\end{figure}

Figure \ref{fig:fig_13}(a) is the spectrum of $\Psi_{u}(t)$, which shows the principal lobe of the single-frequency mode $0.3sin(72\pi t)$  embedded in the spectrum of the sawtooth wave$f_{saw}(t)$. The single-frequency mode $0.3sin(72\pi t)$ is extracted by the orthogonal mode decomposition algorithm so that the sawtooth wave $f_{saw}(t)$ can be accurately restored.
Figure \ref{fig:fig_13}(b) shows the result of the mode decomposition for $\Psi_{u}(t)$.

\section{Conclusion}

Everything in the world has wave properties. Mode decomposition is an important method for understanding the intrinsic characteristics of waves and analyzing the genes of waves. It plays an important role in signal processing, data analysis, machine learning, and other fields.

The previous mode decomposition methods \cite{dragomiretskiy2013variational, huang1998empirical} believed that the modes were narrow band functions, but there were no precise definitions and calculation methods for the bandwidth of the modes. This article provides an accurate definition and computation method for mode bandwidth by calculating the intrinsic phase and instantaneous frequency. Based on this, we proposed the orthogonal projection method for mode extraction. The efficiency of the mode decomposition computation and the unique and orthogonal mode decomposition result are the property of the method. The practical examples showed that good mode decomposition accuracy can be maintained even at the boundary of the time segment, overcoming the 'boundary effects' of previous mode decomposition methods. Of all the approaches of mode decomposition we know in this range, this is the only one with these desirable features.

We introduce the interpolation function $\Psi_{u}(t)$ of finite discrete signals. The interpolation function has the same bandwidth as the finite discrete signal. The interpolation function space $\boldsymbol{IFS_{(n,\Delta)}}$ is constructed. The orthogonal basis and the orthogonal projection operator of $\boldsymbol{IFS_{(n,\Delta)}}$ are found. The modes can be extracted by applying the orthogonal projection operator to $\Psi_{u}(t)$. Based on the parity decomposition of a signal, we define the intrinsic phase function and the instantaneous frequency of the signal. The bandwidth of modes is determined under the constraint that the instantaneous frequency is always positive (or always negative) throughout the time segment. Compared with traditional EMD and VMD algorithms, the orthogonal mode decomposition algorithm has many advantages. It is an algorithm for extracting individual modes and can extract any mode of interest without considering other modes. The parity decomposition of the signals also helps directly obtain the real and imaginary part spectrum. Based on the position of the principal lobe in the real and imaginary part spectrum, the center frequency of the mode can be determined quickly. The bandwidth of the modes can be determined through the iteration procedure, and the iteration procedure of determining the bandwidth of the mode is also the procedure of extracting the mode.

Further experiments show that the orthogonal mode decomposition algorithm is still efficient for the real-time rolling refreshed data set obtained by real-time sampling. Because of high computational efficiency, the orthogonal mode decomposition algorithm is even more suitable for real-time applications. These applications include signal real-time filtering, equipment fault diagnosis, real-time modeling of physical processes, etc.

\end{document}